\documentclass[preprint2]{aastex}
\usepackage{natbib}
\newcommand{\noprint}[1]{}

\shorttitle{Differential rotation within KIC\,3527751}
\shortauthors{Reed et al.}

\begin{document}

\title{The discovery of differential radial rotation in the pulsating
subdwarf B star KIC\,3527751}

\author{H. M. Foster and M. D. Reed}
\affil{Department of Physics, Astronomy, and Materials Science, Missouri
State University, Springfield, MO 65897}

\author{J. H. Telting}
\affil{Nordic Optical Telescope, Rambla Jos\'e Ana Fern\'andez P\'erez 7, 38711 
Bre\~na Baja, Spain}

\author{R. H. \O stensen}
\affil{Instituut voor Sterrenkunde, KU Leuven, Celestijnenlaan 200 D, 3001
Leuven, Belgium}

\author{A. S. Baran}
\affil{Uniwersytet Pedagogiczny, Obserwatorium na Suhorze, ul. Podchor\c{a}\.zych 2, 30-084 Krak\'ow, Poland}

\begin{abstract}
We analyze three years of nearly-continuous 
\emph{Kepler} spacecraft short cadence 
observations of the pulsating subdwarf\,B star KIC\,3527751.
We detect a total of  251 periodicities, most in the $g$-mode
domain, but some where $p$-modes occur, confirming that
KIC\,3527751 is a hybrid pulsator.
We apply seismic tools to the periodicities to characterize
the properties of KIC\,3527751.
Techniques to identify modes include asymptotic period spacing 
relationships, frequency multiplets, and the separation of 
multiplet splittings.
These techniques allow for 189 (75\%) of the 251 periods to be 
associated with pulsation modes. Included in these are three sets of
$\ell\,=\,4$ multiplets and possibly an $\ell\,=\,9$ multiplet.
Period spacing sequences indicate $\ell\,=\,1$ and 2 overtone
spacings of $266.4\,\pm\,0.2\,$ and $153.2\,\pm\,0.2\,$seconds,
respectively. We also
calculate reduced periods, from which we find 
evidence of trapped pulsations. Such mode trappings can be used
 to constrain the core/atmosphere transition layers. Interestingly, 
frequency multiplets in the $g$-mode region,
which sample deep into the star,
indicate a rotation period of $42.6\,\pm\,3.4$\,days while $p$-mode
multiplets, which sample the outer envelope, indicate a rotation period of
$15.3\,\pm\,0.7$\,days. We interpret this as differential rotation in
the radial direction with the core rotating
more slowly. This is the first example of differential rotation for
a subdwarf B star.
\end{abstract}


\section{Introduction}

The \emph{Kepler} spacecraft was launched in 2009 with a primary goal: to discover 
extrasolar planets by means of detecting their transits.  To this end, the 
spacecraft has accomplished its mission; at the time of writing, 
\citet{tjs14} suggest there may be 16\,285 potential stars with transit 
or eclipse detections.

\emph{Kepler's} applications are broader than just hunting for extrasolar planets.  
The spacecraft has also been extremely useful for discovering binary stars, 
which allow us to derive the bulk characteristics of stars, such as mass, 
radius, distance, and luminosity \citep{conroy2014}.  
\emph{Kepler} has also advanced the
field of asteroseismology, or the process of using a star's vibrations to 
determine its physical characteristics. \emph{Kepler's} Earth-trailing orbit
allows it to continuously obtain data, avoiding ground-based limitations,
such as daytime gaps, atmospheric transparency variations, and annual visibility
cycles. \emph{Kepler} only ceased observing during monthly data transmissions and
a few safing events over the course of its four-year program.

Despite asteroseismology being a secondary goal, \emph{Kepler} has been
particularly successful for studying the oscillations of subdwarf B (sdB)
stars.  Subdwarf B stars are 
extreme horizontal branch  stars with  temperatures in the range of
20\,000 to 40\,000\,K.  These stars have shed their 
outer layers near the 
tip of the red giant branch and have become the exposed cores of horizontal 
branch stars \citep[for a review of 
the properties of sdB stars see][]{heber09}.  
The pulsations of sdB stars are divided into two categories
 based on their periods.  Short period pressure ($p$-)mode pulsators, 
or V361\,Hya stars, have amplitudes 
typically less than 1\% of their mean brightness 
and pulsation periods of just a few minutes.  Long period gravity 
($g$-)mode pulsators are classified as V1093\,Her stars with typical
amplitudes below $0.1\%$ and longer pulsation periods typically near
an hour.  
Some stars exhibit both kinds of pulsation.  These 
hybrids usually exhibit one pulsation type more strongly than the other, 
with one exception; KIC\,9472174 which shows an abundance of both types of 
pulsation modes \citep{2m1938}. The two classes are also separated in
temperature with the V1093\,Her stars being
cooler than the  V361\,Hya stars
\citep[for a review of sdB pulsation properties, see][]{ostensen10}. In
this paper we use the term sdBV to generically indicate pulsating sdB
stars.

\emph{Kepler's} nearly-continuous observations
have been particularly useful for 
the study of the longer period $g$-mode pulsators, for
which ground-based data have not provided seismic solutions.
At the time of writing, 17 papers have been published using \emph{Kepler} 
observations of sdBV stars.
Previously published results using \emph{Kepler} data include
the discovery of nearly-evenly-spaced $g$-mode periods \citep{reed11c} and
the detection of frequency 
multiplets \citep{baran12a,telting12a,ostensen12b,ostensen14a,reed14}, 
both of which can be used to identify pulsation modes. 
Asteroseismic rotation periods have 
been found to be on the order of tens to a hundred days. This includes
binary stars which have been shown to be subsynchronous rotators, even with
orbital periods as short as half a day \citep{pablo11,pablo12,
telting12a,jht14,ostensen14a}.

The objective of this or any other asteroseismological study is to 
characterize the physical properties of the star in question.  Model
constraints include spectroscopic measurements ($T_{\rm eff}$ and $\log g$),
as well as asteroseismic properties. Asteroseismic constraints can be as
simple as a list of frequencies to compare with models 
\citep[as in the papers by][using survey-phase \emph{Kepler} data]{vvg10,
charpinet11a},
or quite detailed, as in small structures in Echelle diagrams. 
Tools which have been applied to sdBV stars with some success include
measuring stellar rotation via frequency multiplets \citep[e.g.][]{telting12a};
 period spacings for mode identification \citep[e.g.][]{reed11c}; 
long overtone sequences observed in Echelle diagrams \citep[e.g.][]{reed14}, 
as well as small deviations \citep[e.g.][]{baran12c}; regions above and below 
which period spacings
do not behave asymptotically; reduced period diagrams for detecting trapped
modes \citep[][]{ostensen14a}; and sliding Fourier Transforms (sFTs) 
to resolve amplitude variations \citep[][]{telting12a,reed14}.

The target of this paper, KIC\,3527751, was examined in a preliminary study
of hybrid sdB pulsators by \citet{reed10a}. They detected 
41 $g$- and 3 $p$-mode periodicities from
one-month of \emph{Kepler} survey data. 34
of the $g$-modes were identified as $\ell\,=\,1$ or 2 
using period spacings.  
Subsequent to \emph{Kepler's} survey phase, KIC\,3527751 was continuously 
observed with one minute cadence
from quarter 5 (Q5) until mission end during Q17. 
Here we analyze all available data (1148\,days), 
providing our best estimate of the 
frequency content to which we apply our asteroseismological tools,
as discussed above.

\begin{figure*}[!htbp]
\figurenum{1}
\includegraphics[angle=-90,width=5.0 in]{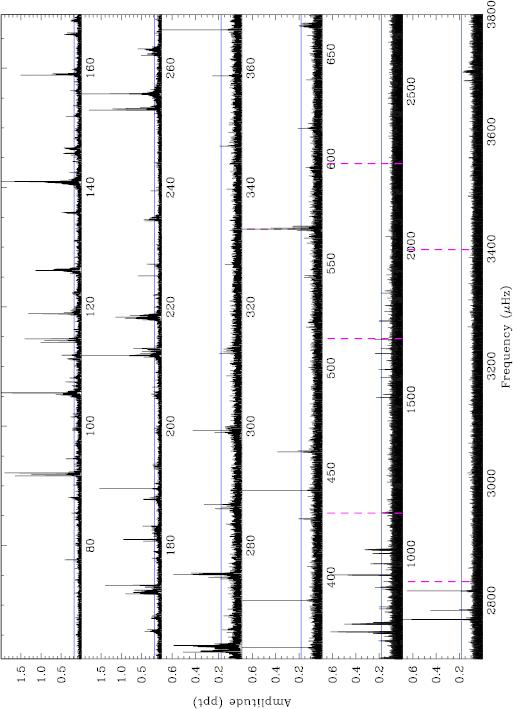}
\caption{Fourier transform of KIC\,3527751. The solid (blue) horizontal
line is the $5\sigma$ detection limit and the dashed (magenta)
vertical lines indicate known frequencies where spacecraft artifacts may occur.}
\label{fig01}
\end{figure*}
\notetoeditor{color figure in online edition}

\section{Data Analysis}

\emph{Kepler} obtains data in two modes: short cadence (SC), which
produces one integration every 58.85 seconds, and long cadence (LC), which
produces an integration every 30 minutes.  Short candence observations
consume more
of \emph{Kepler's} limited memory, so the targets observed in this mode 
are far fewer.
We downloaded 
optimally-extracted lightcurves from the Mikulski Archive for Space
Telescopes\footnote{http://archive.stsci.edu/kepler/},
removed long-term trends ($>1.5$ days) with low-order spline fitting, and 
normalized the data by mean brightness.  We sigma-clipped the data at 
$5\sigma$ and multiplied the modulation intensities so amplitudes would 
appear in the Fourier transform (FT) as parts-per-thousand (ppt).

The data span almost 1148 days and include 1.55 million data 
points which is a temporal resolution of $0.010\,\mu$Hz.  
These data have a duty 
cycle of 92.4\%, with the largest gaps caused by spacecraft safing events 
during Q8, Q14, and Q16.  As noted by \citet{roy14b}, a $4\sigma$ detection
limit would contain $\sim50$ spurious peaks, so
we increased our detection limit to $5\sigma$, to be reasonably confident
that detections are real signal.  
 We determined the mean level
in the FT to be $\sigma_{FT}\,=\,0.032\,ppt$, making our $5\sigma$ limit
0.16\,ppt.
It should be noted that a $5\sigma$ detection limit only applies when
examining the entire FT. When looking for predicted periodicites in 
smaller regions (of, say a few hundred $\mu$Hz), a $4\sigma$ detection limit
is perfectly justified.
Significant 
peaks range from 72 to $3704\,\mu$Hz, though they are concentrated between
$100-300\,\mu$Hz.  Peaks occurring below $20\,\mu$Hz are attributed to 
residual signal from the spacecraft which was not removed by our detrending.
An FT of the data is shown in Fig.\,\ref{fig01} with the
detection limit indicated by a horizontal line.

As discussed in \citet{reed14}, the previously utilized method of 
fitting the original lightcurve with a non-linear least-squares program 
and prewhitening the resulting Fourier transform (FT) would have been not 
only tedious, but rather counterproductive as exceptionally few peaks could 
be cleanly removed.
 We attribute this property to the long-term instability 
of amplitudes and even, occasionally, of frequencies. In order to evaluate
the pulsation content, we used the two tools described in \citet{reed14}:
sliding FTs (sFTs) to evaluate the pulsation content in the
time domain and Lorentzian peak fitting which serves as a method for
determining peak widths as an indicator of frequency errors.

\begin{figure*}[!htbp]
\figurenum{2}
\epsscale{2.0}
\plottwo{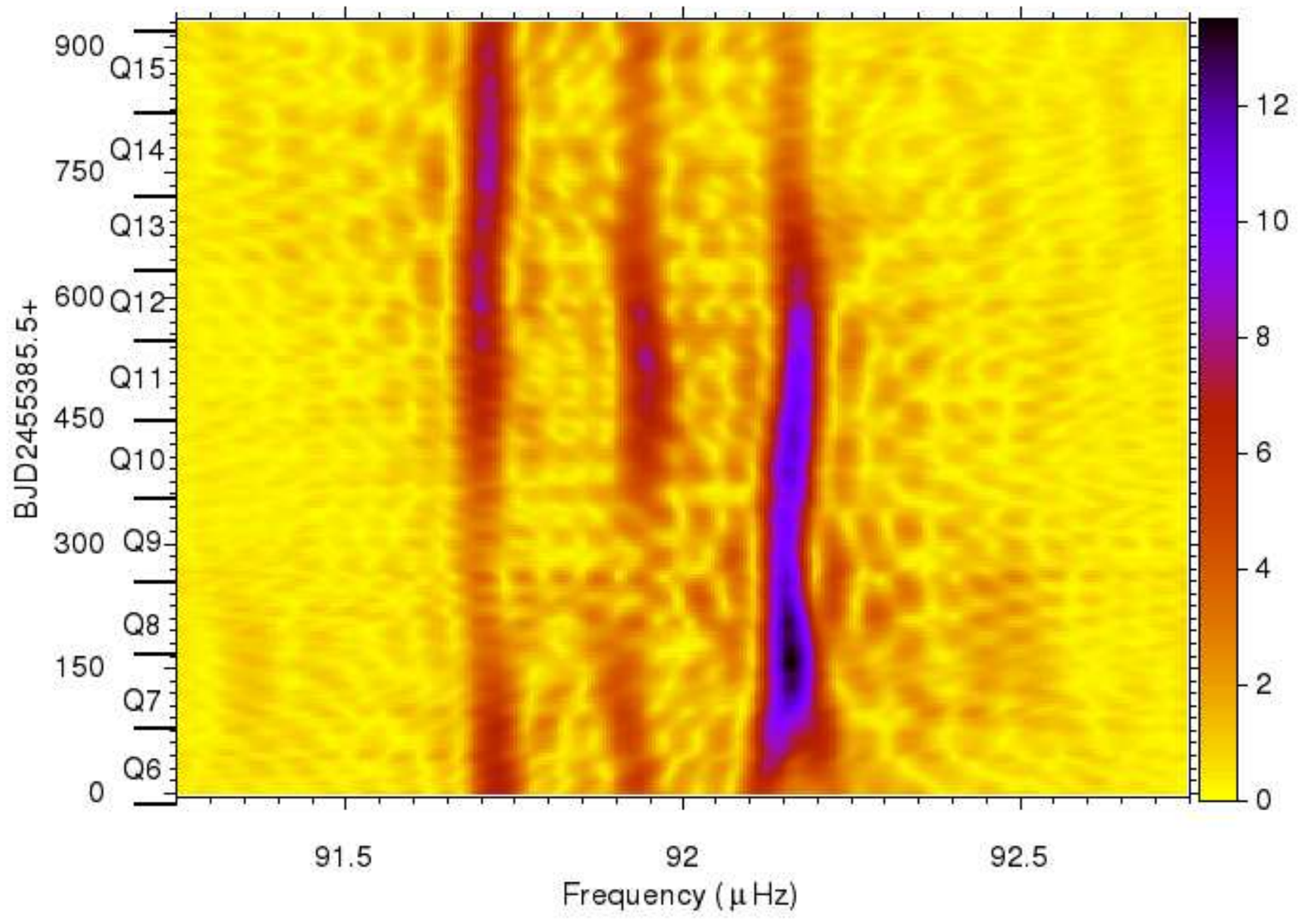}{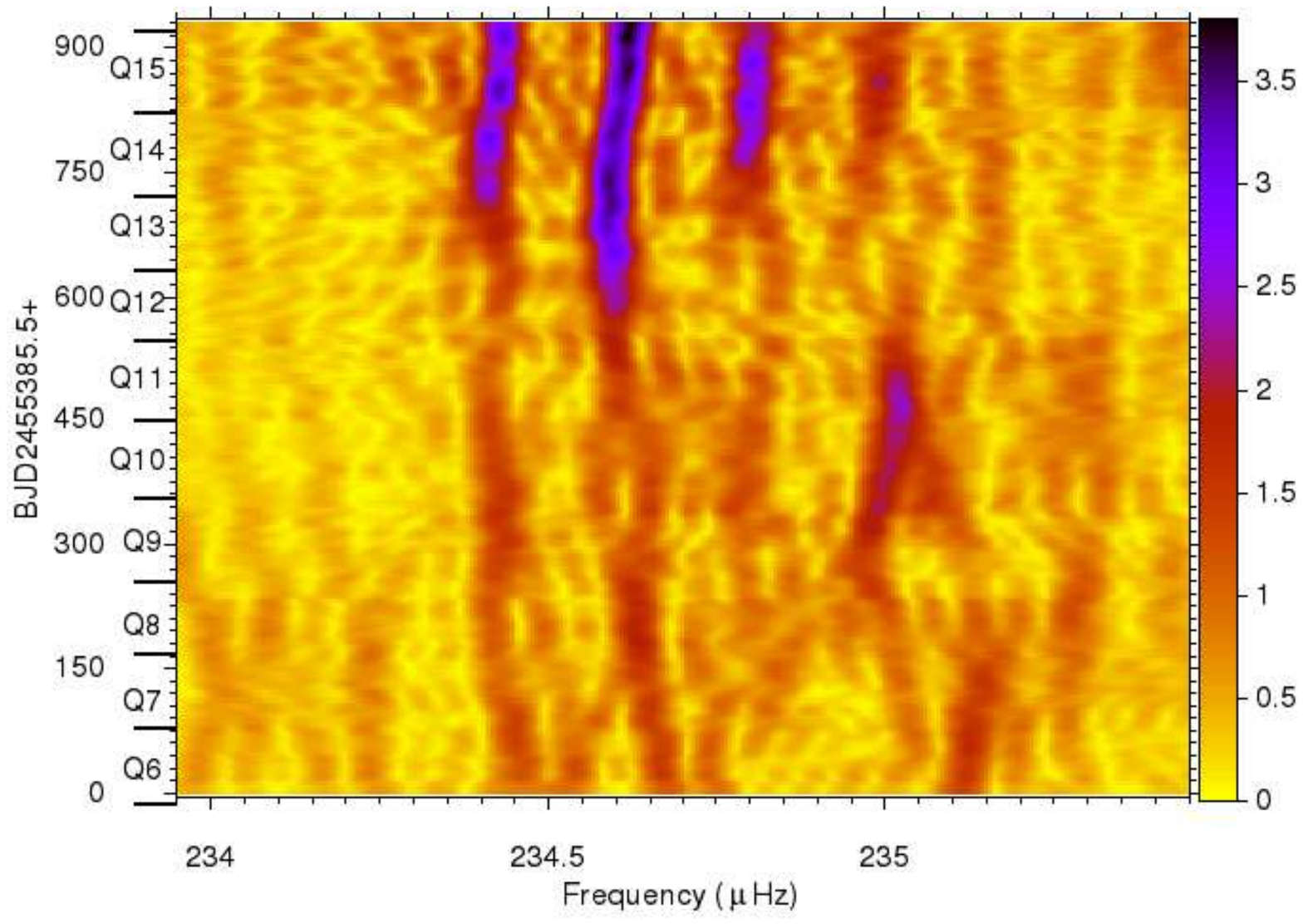}
\caption{Examples of sliding Fourier transforms (sFTs). A complete set
is available in the on-line material.}
\label{fig02}
\end{figure*}
\notetoeditor{color figure in online edition}

Frequency multiplets are readily apparent in the FT (see Fig.\,\ref{fig03})
of the complete data set and were used as a guide for the sFTs. Sliding
FTs were generated using data spanning 220\,days, to fully resolve
freqency multiplets, and stepped by 5 days through the entire data set.
Sample sFTs are shown in Fig.\,\ref{fig02} with sFTs
of the entire frequency spectrum available on-line.
The sFTs were used as
guides for the Lorentzian fitting. If amplitudes were only detectable
during a portion of the data, Lorentzian fits were obtained from only
those portions (using data spanning a minimum of 200 days to insure 
frequencies are resolved). In total, we fitted 251 frequencies, which
were compared with known spacecraft artifacts to ensure none are in
our list (except $f159$ as noted in \S 3.1). 
Table\,\ref{tab01} provides the pulsation frequencies, Lorentzian widths,
and amplitudes resulting from our FT fitting.

\begin{deluxetable}{lcccccccccc}
\tablecaption{Table of asteroseismic quantities. Column\,1 provides a
label for the periodicity, Columns\,2 and 3 the frequency and period,
with errors in parentheses, and column\,4 the observed amplitude 
 taken from the Lorentzian fit of the entire data set. Columns\,5
through 8 provide our best estimate mode identifications, Columns\,9
and 10 period spacing deviations, and Column\,11 the frequency splitting
(from the subsequent frequency) of multiplet members. $^{\dag}$ indicates
periodicities detected by \citet{reed10a} and $^*$ indicates frequencies which
are listed in Table\,\ref{combo} as part of a combination frequency. 
Table\,1 is published in its entirety in the electronic
edition of the Astrophysical Journal. A portion is shown here for
guidance regarding its form and content.}
\tablehead{
\colhead{ID} & \colhead{Freq.} & \colhead{Period} & \colhead{Amp.} &
\colhead{$\ell$} & \colhead{$m$} & \colhead{$n_{\ell\,=\,1}$} & \colhead{$n_{\ell\,=\,2}$} &
\colhead{$\left(\frac{\Delta P}{P}\right)_{\ell\,=\,1}$} & \colhead{$\left(\frac{\Delta P}{P}\right)_{\ell\,=\,2}$} & \colhead{$\delta f$}\\
& \colhead{$\mu$Hz} & \colhead{sec.} & \colhead{ppt} &\nodata &\nodata &\nodata &\nodata &\nodata &\nodata & \colhead{$\mu$Hz} }
\startdata
f010 & 91.703 (0.008) & 10904.73 (0.91)& 1.64 & 2 & -1 & -- & 68 & -- & 0.35 & 0.231 \\
f011 & 91.934 (0.007) & 10877.33 (0.85)& 1.24 & 2 & 0 & -- & 68 & -- & 0.18 & 0.2266 \\
f012$^{\dag}$ & 92.161 (0.010) & 10850.59 (1.21)& 1.89 & 2 & 1 & 38 & 68 & 0.07 & 0.00 & --  \\
f013 & 94.590 (0.032) & 10571.94 (3.56)& 0.38 & 1 & -- & 37 & -- & 0.02 & -- & --  \\
f014 & 99.289 (0.004) & 10071.59 (0.41)& 0.22 & 2 & 0 & -- & 63 & -- & -0.08 & 0.1221 \\
f015$^*$ & 99.411 (0.005)& 10059.22 (0.47)& 0.19 & 1 & 1 & 35 & 63 & 0.09 & -0.16 & 0.1141 \\
\hline
\enddata
\label{tab01}
\end{deluxetable}

\subsection{Combination frequencies}
A search for combination frequencies and a likelihood
comparison was completed as in \citet{telting12a}. Residuals were calculated
using the form $\delta f\,=\,f_3-f_2-f_1$ for all combinations of frequencies
and those with $\delta f\,=\,0$ within the errors were reported for
examination. Eight combinations were discovered having $\delta f\,<\,0.0001\,
\mu$Hz. Of these, three only involve low-amplitude frequencies which are
unlikely to produce combination modes, 
while five have at least one higher-amplitude
frequency. We consider it likely that these five actually are combination
frequencies and they are listed in 
Table\,\ref{combo} and marked in Table\,\ref{tab01}
with asterisks.

\begin{deluxetable}{cccccc}
\tablecaption{Possible combination frequencies  (in $\mu$Hz) 
for $f_3-f_2-f_1\,=\,
\delta f$ with $\delta f\,<\,0.0001$. Amplitudes  (in ppt) 
are provided in Columns
4 -- 6.}
\tablehead{
\colhead{$f_1$} & \colhead{$f_2$} & \colhead{$f_3$} &
\colhead{$A_1$} & \colhead{$A_2$} & \colhead{$A_3$} }
\startdata
76.174 & 135.710 & 211.884 & 0.16 & 0.35 & 3.30\\
99.411 & 118.866 & 218.277 & 0.19 & 1.31 & 1.14 \\
105.632 & 193.670 & 299.302 & 2.92 & 0.22 & 0.42 \\
158.845 & 774.149 & 932.994 & 1.51 & 0.22 & 0.62\\
253.194 & 312.699 & 565.893 & 1.21 & 0.13 & 0.21 \\
\hline
\enddata
\label{combo}
\end{deluxetable}

\subsection{Spectroscopy}

Over the 2010 and 2011 observing seasons of the \emph{Kepler} field, we
obtained a total of 24 spectra of KIC\,3527751. 
Low-resolution spectra (R\,$\approx$\,2000--2500) have been collected 
using the Kitt Peak 4-m Mayall telescope with RC-Spec/F3KB, the kpc-22b 
grating and a 1.5--2.0\,arcsec slit, the 2.56-m Nordic Optical Telescope 
with ALFOSC, grism \#16 and a 0.5\,arcsec slit, and the 4.2-m William 
Herschel Telescope with ISIS, the R600B grating and 0.8--1.0\,arcsec slit.  
Exposure times were 600\,s at KP4m and WHT, and either 500\,s or 300\,s at 
the NOT. The resulting resolutions based on the width of arc lines is 
1.7\,\AA\ for the KP4m and WHT setups, and 2.2\,\AA\ for the setup at the 
NOT. See Table\,\ref{tab02} for an observing log.

\begin{deluxetable}{lcccccc}
\tablecaption{Log of the low-resolution spectroscopy of KIC\,3527751. }
\tablehead{
\colhead{Mid-exposure date} & \colhead{Barycentric JD} & \colhead{S/N} &
\colhead {RV} & \colhead{RV$_{\rm err}$} & \colhead{Telescope} &
\colhead{Observer} \\
\colhead{} & \colhead{$-2\,455\,000$} & \colhead{}  & \colhead{km\,s$^{-1}$} & 
\colhead{km\,s$^{-1}$} &\colhead{} & \colhead{} }
\startdata
2010-07-28\,01:32:16 & 405.5669179 & 107.4 & 6.9 & 3.1 & WHT  & RH\O \\
2010-07-28\,05:02:45 & 405.7130811 & 82.7 & -6.0 & 3.1 & WHT  & RH\O \\
2010-08-13\,08:25:22 & 421.8535618 & 69.8 & -1.2 & 6.1 & KPNO & MDR/LHF \\
2010-08-13\,08:36:39 & 421.8613996 & 65.3 & -1.5 & 9.6 & KPNO & MDR/LHF \\
2010-08-14\,07:11:56 & 422.8025448 & 57.4 & 5.2 & 4.1 & KPNO & MDR/LHF \\
2010-08-14\,07:22:55 & 422.8101739 & 63.8 & 11.6 & 6.5 & KPNO & MDR/LHF \\
2010-08-14\,07:33:15 & 422.8173514 & 65.7 & 15.5 & 10.4 & KPNO & MDR/LHF \\
2011-06-01\,02:47:19 & 713.6181963 & 71.7 & -13.0 & 6.1 & NOT  & JHT \\
2011-06-07\,04:11:17 & 719.6767124 & 47.4 & -5.1 & 6.3 & NOT  & JHT \\
2011-06-09\,05:15:40 & 721.7214831 & 42.8 & -15.3 & 6.3 & NOT  & JHT \\
2011-06-20\,00:33:42 & 732.5259579 & 51.9 & -11.0 & 5.1 & NOT  & JHT \\
2011-06-20\,02:43:12 & 732.6158815 & 48.6 & -5.8 & 5.9 & NOT  & JHT \\
2011-06-20\,05:22:41 & 732.7266388 & 36.2 & -5.9 & 5.7 & NOT  & JHT \\
2011-06-27\,22:53:21 & 740.4564180 & 42.8 & -35.0 & 8.5 & NOT  & JHT \\
2011-07-23\,00:58:51 & 765.5437440 & 43.6 & -26.9 & 9.8 & NOT  & JHT \\
2011-07-23\,01:38:08 & 765.5710230 & 38.5 & -29.0 & 11.2 & NOT  & JHT \\
2011-07-23\,02:18:10 & 765.5988233 & 38.3 & -13.4 & 7.5 & NOT  & JHT \\
2011-07-23\,03:11:48 & 765.6360675 & 47.9 & -37.7 & 9.8 & NOT  & JHT \\
2011-07-23\,04:05:02 & 765.6730338 & 43.2 & -34.8 & 11.2 & NOT  & JHT \\
2011-07-23\,05:00:47 & 765.7117481 & 42.9 & -33.0 & 13.4 & NOT  & JHT \\
2011-07-24\,04:51:35 & 766.7053592 & 47.8 & -17.0 & 9.7 & NOT  & JHT \\
2011-08-29\,00:08:25 & 802.5080643 & 31.8 & -76.3 & 10.4 & NOT  & JHT \\
2011-08-30\,00:30:11 & 803.5231446 & 42.1 & -24.7 & 10.4 & NOT  & JHT \\
2011-08-30\,20:53:13 & 804.3724475 & 40.0 & -23.2 & 10.9 & NOT  & JHT \\
\hline
\enddata
\label{tab02}
\end{deluxetable}

The data were homogeneously reduced and analysed.  Standard reduction 
steps within IRAF include bias subtraction, removal of pixel-to-pixel 
sensitivity variations, optimal spectral extraction, and wavelength 
calibration based on arc-lamp spectra. To fully account for the blue CCD 
etching pattern in the NOT spectra, spectroscopic flats were constructed 
by interpolating between UBV imaging flats along the dispersion direction, 
as halogen flats suffered from stray light in the blue part of the 
spectrum. The target spectra and the mid-exposure times were shifted to 
the barycentric frame of the solar system.  The spectra were normalised to 
place the continuum at unity by comparing with a model spectrum for a star 
with similar physical parameters as we find for the target (described
below).

Radial velocities were derived with the FXCOR package in IRAF. We used
the H$\gamma$, H$\delta$, H$\zeta$ and H$\eta$ lines to determine the
radial velocities (RVs), and used the spectral model fit (described below) 
as a template.  See Table~\ref{tab02} for the results,
with errors in the radial velocities as reported by FXCOR.  The errors
reported by FXCOR are correct relative to each other, but may need
scaling depending on, amongst other things, the parameter settings and
the validity of the template as a model of the star.  

We find that the RV data has more scatter than one would assume for a 
non-variable star, but that the RV data are not sufficient to constrain
any orbital parameters.  Excluding the datapoint from the worst S/N 
spectrum, the average RV$\,=\,-13.0\pm\,3.2\,km\,s^{-1}$, with an RMS 
scatter of $15.1\,km\,s^{-1}$ which is much larger than the errors in the data 
that range from 3--13$\,km\,s^{-1}$, with median individual error of 
$7.8\,km\,s^{-1}$.
An FT gives maximum amplitude of $22\,km\,s^{-1}$ in the frequency range of 
0--17.4\,$\mu$Hz, and lower amplitudes at higher frequencies.

As in our previous papers \citep[e.g.][]{ostensen10b}, 
we have fitted the average spectrum from each observatory
to model grids, in order to determine effective temperature
($T_{\rm eff}$), surface gravity ($\log g$), and photospheric helium abundance
($\log y\,=\,\log N_{\rm He}/N_{\rm H}$). The fitting
procedure used was the same as
that of \citet{edel03}, using the metal-line blanketed LTE
models of solar composition described in \citet{heber99b}.
Mean values from these three fits were
computed using the formal fitting errors as weights and systematics
between observatories were then factored into the errors. The resultant
measurements are indicated in Table\,\ref{tab03} and we adopt the
values of $T_{\rm eff}\,=\,27818\,\pm\,163$, $\log g\,=\,5.35\,\pm\,0.03$,
and $\log \left(N_{He}/N_H\right)\,=\,-2.99\,\pm\,0.04$.

\begin{deluxetable}{llll}
\tablecaption{Spectroscopic parameters from the detrended mean spectra. }
\tablehead{
\colhead{Telescope} & \colhead{$T_{\rm eff}$} & \colhead{$\log g$} & 
\colhead{$\log\left(N_{\rm He}/N_{\rm H}\right)$} \\
 & \colhead{[K]} & \colhead{[dex]} & \colhead{[dex]} }
\startdata
KPNO & $27945\,\pm\,133$ & $5.35\,\pm\,0.02$ & $-2.97\,\pm\,0.03$ \\
NOT1 & $27715\,\pm\,102$ & $5.32\,\pm\,0.02$ & $-3.01\,\pm\,0.03$ \\
NOT2 & $27709\,\pm\,102$ & $5.33\,\pm\,0.02$ & $-3.02\,\pm\,0.02$ \\
WHT & $27913\,\pm\,82 $ & $5.38\,\pm\,0.01$ & $-2.95\,\pm\,0.03$ \\ \hline
Adopted & $27818\,\pm\,163$ & $5.35\,\pm\,0.03$ & $-2.99\,\pm\,0.04$ \\
\hline
\enddata
\label{tab03}
\end{deluxetable}

\section{Mode Identification}

Prior to \emph{Kepler}, observational mode identifications were extremely
rare for sdB pulsators, leaving period matching between models and 
observations (the forward method) as virtually the only means of correlating
periodicities to modes for sdB stars.  However, using \emph{Kepler} extended
data sets, purely observational mode identifications using frequency
multiplets and period spacings have become well-established techniques.
We follow the examples of \citet{baran12a} and \citet{reed11c,reed14} 
in applying
these methods to identify modes in KIC\,3527751.  To first order, stellar 
rotation removes azimuthal frequency degeneracy resulting in frequency
multiplets appearing with $2\ell\,+\,1$ members, with each member
shifted by 
\begin{equation}\Delta\nu\,=\,\Delta m\Omega\left(1\,-\,C_{n,\ell}\right)\end{equation}
from the $m\,=\,0$ value \citep{led51}. 
$\Omega$ is the frequency of stellar rotation,
and $C_{n,\ell}$ is the Ledoux constant which is nearly zero for $p$-modes
and for $g$-modes depends on the
mode degree as 
\begin{equation}C_{n,\ell}\,\approx\,\frac{1}{\ell\left(\ell\,+\,1\right)}.\end{equation}
Similarly, $g$-modes may show radial overtones spaced evenly in
\emph{period} as 
\begin{equation}\Delta \Pi_{\ell}\,=\,\frac{\Pi_o}{\sqrt{\ell\left(\ell +1\right)}}
\end{equation}
where $\Delta \Pi_{\ell}\,=\,\Pi_{\ell ,n+1}-\Pi_{\ell ,n}$ and
$\Pi_o$ is the radial fundamental period. Even period spacings imply
idealized homogenous stars using asymptotic relations ($n\gg\ell$). In
practice, even period spacings have \emph{not} been observed at short
(where $n$ is not much greater than $\ell$) or long (where frequency
multiplets become spaced similar to period spacings) periods
\citep[][]{baran12c,ostensen14a,reed14}.

\subsection{Frequency Multiplets}
\paragraph{$g$-mode multiplets:}
Uncovering multiplets in KIC\,3527751 is extremely useful in associating 
periodicities with pulsation modes, $\ell$ and $m$.  A cursory inspection of 
the most visible multiplets in the $g$-mode region 
reveals a common splitting near $0.23\,\mu$Hz.  
Many of these multiplets have four or five peaks, meaning these are 
most likely $\ell\,=\,2$ modes. (A discussion of how inclination
affects the number of observed multiplet members is in \S 4.4)
22 multiplets with similar splittings were found and designated as
$\ell\,=\,2$ modes.  In cases where 
the $m\,=\,0$ component was ambiguous, we arbitrarily assigned it
to the appropriate highest amplitude  member.  The average
frequency splitting of the most easily distinguished $\ell\,=\,2$
 multiplets was $0.23\,\pm\,0.03\,\mu$Hz.  
Triplets which displayed similar splittings were likewise designated 
as $\ell\,=\,2$.

Though the overwhelming majority of multiplets were designated $\ell\,=\,2$, 
some had splittings smaller than $0.23\,\mu$Hz, often closer to $0.13\,\mu$Hz.
This value is the splitting we expect to see from Eqns.\,1 and 2 for
$\ell\,=\,1$ multiplets.
Nine multiplets 
had splittings near this value and were designated as $\ell\,=\,1$.  We find 
the average splitting of $\ell\,=\,1$ multiplets to be $0.12\,\pm\,0.03\,\mu$Hz.
  Altogether, we associated 123 frequencies with $\ell\,=\,1$ and $2$ modes
using the frequency multiplet splitting method.

\begin{figure}[!htbp]
\figurenum{3}
\epsscale{0.99}
\plotone{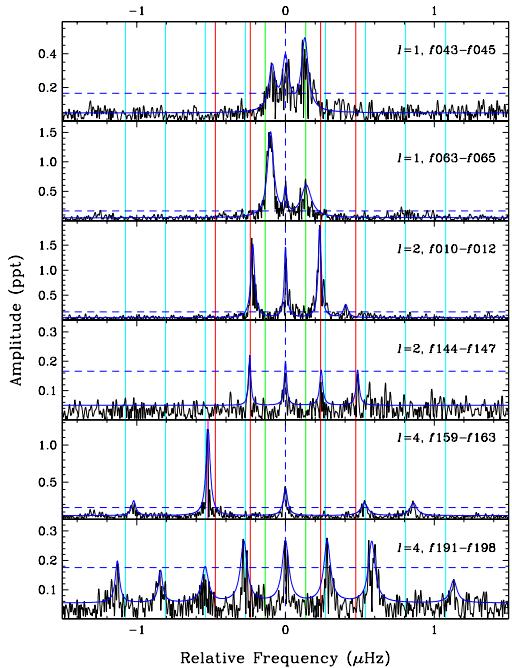}
\caption{Sample frequency multiplets and Lorentzian fits
for $\ell\,=\,1$, 2, and 4 modes.
Frequency is relative to the center of the multiplet (dashed vertical
blue line) with other vertical lines indicating the
$\ell\,=\,1$ (green), 2 (red), and 4 (cyan) multiplet spacings. The
wide solid vertical (magenta) line indicates the 1/LC spacecraft artifact.
The horizontal (blue) dashed line is the $5\sigma$ detection limit. Not all
fitted frequencies have amplitudes that are significant in the full data set,
but are during subsets.}
\label{fig03}
\end{figure}
\notetoeditor{color figure in online edition}

One multiplet present in KIC\,3527751 ($f158$ - $f162$) at first glance 
appears to be a quintuplet, but the spacing between members is about
double that expected for $\ell\,=\,2$ multiplets. Additionally,
there are two multiplets ($f164$ - $f169$ and $f190$ - $f197$)
which contain too many members 
(six and eight) to be $\ell\,=\,2$ modes and the splitting
is slightly wider, consistent with the smaller Ledoux constant 
expected for
$\ell\,=\,4$ $g$-modes. In total, 19 frequencies are associated with
these three multiplets, with an average $\Delta m\,=\,1$ splitting of 
$0.27\,\pm\,0.04\,\mu$Hz.
We interpret these three multiplets to be $\ell\,=\,4$ modes. 
Sample multiplets with
their Lorentzian fits are shown in Fig.\,\ref{fig03}.

We note that
$f159$ is at the known 1/LC spacecraft artifact. Typically the
1/LC is \emph{not} observed, but rather 8-10/LC more commonly are,
yet this gave us pause. We examined the nearest star
with SC data (KIC\,3323887), processing those data in the same way as for
KIC\,3527751, to see if the 1/LC artifact appears. There was no signal
at 1/LC in KIC\,3323887. We also compared the sFT for $f159$
with other LC artifacts at $4531$ (8/LC) and $5098\,\mu$Hz (9/LC), 
which show a harmonic frequency shift, since intrinsic frequencies
become Doppler shifted due to the spacecraft's orbit around the Sun
after the heliocentric correction is applied.
The sFT for $f159$
does not appear similar to those for 8/LC or 9/LC and so we suggest 
that $f159$ is intrinsic to
KIC\,3527751.

\begin{figure*}[!htbp]
\figurenum{4}
\includegraphics[angle=-90,width=5.0in]{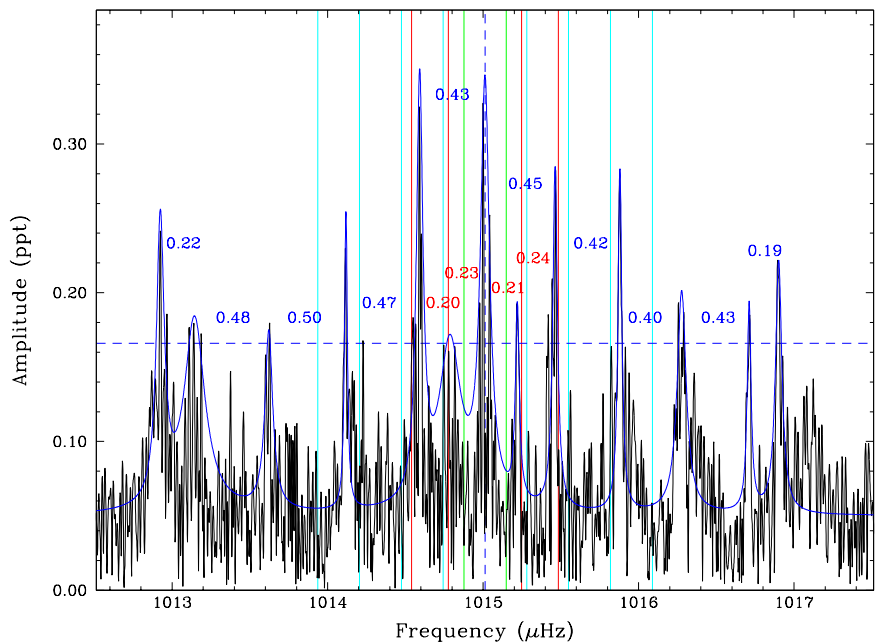}
\caption{Possible high-degree multiplet. The solid black line is the 
FT from the entire data set with the blue lines the Lorentzian fits, as
in Fig.\,\ref{fig03} and the horizontal (blue) dashed line is the $5\sigma$ 
detection limit. The vertical lines indicate the central frequency (dashed)
and the $\ell\,=\,1$ (green), 2 (red), and 4 (cyan) multiplet spacings
from Fig.\,\ref{fig03}. The numbers indicate the frequency spacings, in
$\mu$Hz.}
\label{fthd}
\end{figure*}

Another interesting multiplet includes $f213$ -- $f225$ which is shown
in Fig.\,\ref{fthd}. Eight frequencies are
spaced by 0.40 to 0.50$\mu$Hz or 19 frequencies are spaced by 0.19 to 
0.25$\mu$Hz missing members corresponding to $m\,=\,\pm3$, 5, and 7
(with the $m\,=\,-1$ frequency listed as \emph{tentative} in Table\,1
since it is just below $5\sigma$).
If we assume the smaller spacing,
then the average $\Delta m\,=\,1$
frequency splitting would be $0.22\,\pm\,0.02\,\mu$Hz and
the mode would be $\ell\,\geq\,9$. However, the splittings would be
slightly smaller than the $\ell\,=\,2$ ones, which is contary
to Eqn.\,2. On 
the other hand, if we assume the larger frequency splitting (with the
outside pair just being chance alignments),
then it would be an  $\ell\,\geq\,4$ multiplet, but
with spacings of $0.45\,\pm\,0.03\,\mu$Hz which would be 67\% larger
than the other $\ell\,=\,4$ multiplets. 
We tentatively assign it to the 
smaller splittings as an $\ell\,=\,9$ multiplet in Table\,\ref{tab01} with
the caveat that low and variable amplitude signals are difficult to
extract from the FT and may be imprecise.

\notetoeditor{color figure in online edition}

\paragraph{$p$-mode multiplets:}
Looking beyond about six minutes, where we expect $p$-mode pulsations to occur,
there are 10 frequencies with amplitudes above the $5\sigma$ limit, not including known
artifacts. These fall into sets near 2780 and $3700\,\mu$Hz. Additionally, the
set near $2780\,\mu$Hz has  four frequencies with amplitudes above
the $4\sigma$ limit which perfectly fit into two sets of frequency multiplets
($f239$--$f241$ and $f244$--$f245$).

These 14 frequencies bear some resemblance
to Balloon\,090100001 \citep{baran09}. The highest-amplitude frequency
($f238$) occurs roughly where the radial fundamental is typically
found in such stars,
followed by a much lower amplitude triplet ($f239$ - $f241$) and 
quintuplet ($t242$ - $f246$), in order of
ascending frequency. These multiplets have consistent frequency splittings
of 0.7 to $0.8\,\mu$Hz. Beyond this group is another group near
$3700\,\mu$Hz with inconsistent frequency splittings, also similar to
what is observed for Balloon\,090100001. The ratio of the $f238$--$f246$
group near $2\,800\,\mu$Hz to the $f247$--$f251$ group near
$3\,700\,\mu$Hz is 0.76 which is similar to, but slightly
larger than, the value of 0.7
calculated for Balloon\,090100001 \citep{baran09}.

Based on the consistent frequency splittings
and pattern similarities with Balloon\,090100001, we assign the first
nine $p$-mode frequencies as an $\ell\,=\,0$ singlet, an $\ell\,=\,1$
triplet and an $\ell\,=\,2$ quintuplet. The remaining five
frequencies are most likely the next overtone sequence, but we cannot
assign specific modes to frequencies as their splittings are inconsistent.

\subsection{Period spacings}

\begin{figure*}[!htbp]
\figurenum{5}
\includegraphics[angle=-90,width=5.0in]{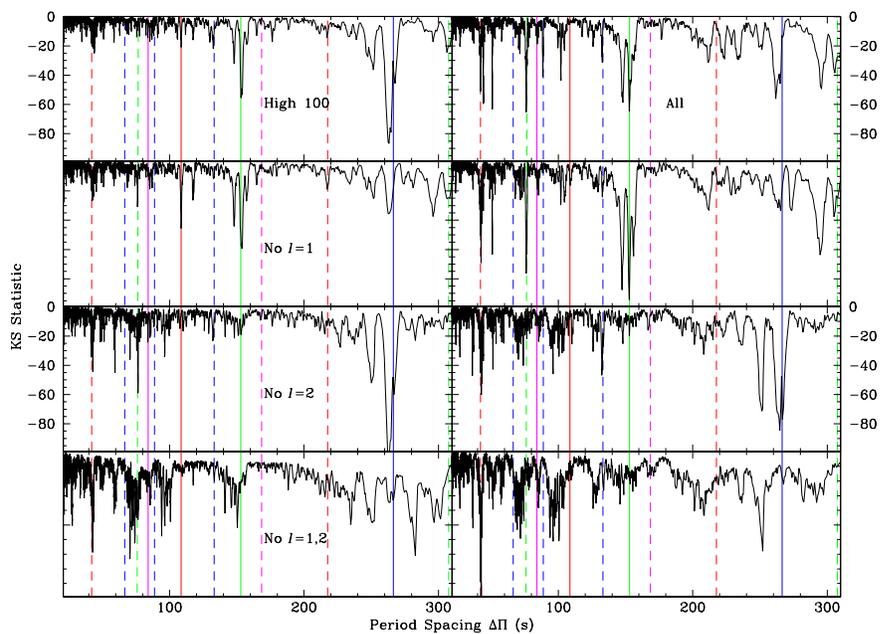}
\caption{KS diagram indicating period spacing sequences. The left panels only
include the highest 100 amplitude periods while the right panels include
all periods longer than 900\,seconds. Solid lines near 153 and 266\,seconds indicate
the $\ell\,=\,2$ and 1 sequences, respectively, and the solid lines near 84 and 109
seconds indicate where the $\ell\,=\,4$ and 3 sequences should appear. Dashed lines
indicate sequence aliases. (Color-coded version appears on-line.)}
\label{fig04}
\end{figure*}

To begin the hunt for period spacing sequences in KIC\,3527751, we 
applied a Kolmogorov-Smirnov (KS) test, using the periods listed in 
Table\,\ref{tab01}.  However, all 251 periods do not produce any
distinguishing peaks.
Since geometric cancellation increases with increasing
$\ell$ \citep{pes85}, we did a KS test on just the 100 highest-amplitude
periods. This produced a substantial peak at 265\,seconds
with a lesser one at 153\,seconds, right where
expected from Eqn.\,3 ($265/\sqrt{3}=153\,$seconds). Both KS tests are
shown in the top panels of Fig.\,\ref{fig04}.

\notetoeditor{color figure in online edition}

\begin{figure*}[!htbp]
\figurenum{6}
\includegraphics[angle=-90,width=5.0in]{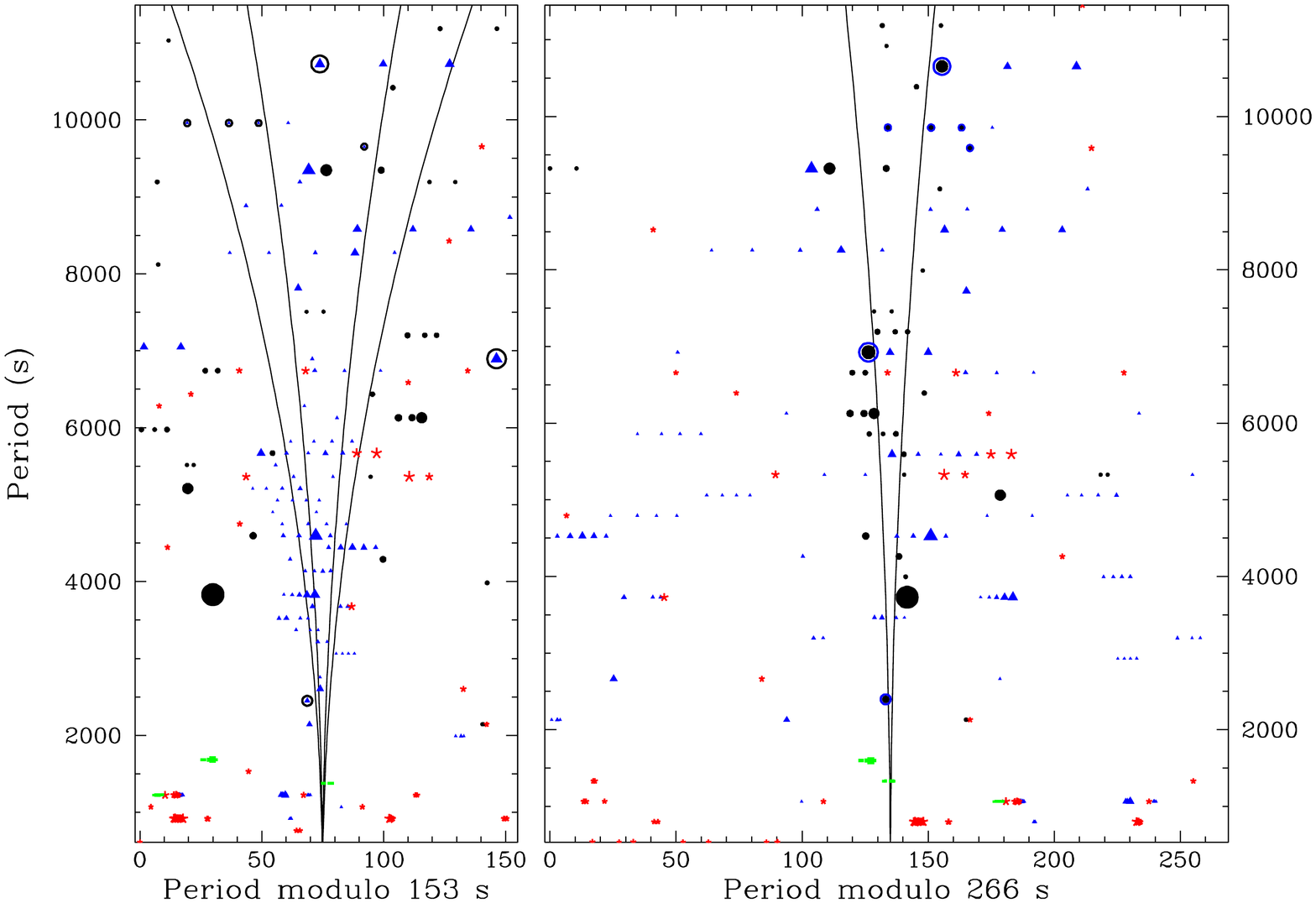}
\caption{Echelle diagrams for $\ell\,=\,1$ (right) and 2 (left)
modes. Black circles indicate our identified $\ell\,=\,1$ modes,
blue triangles $\ell\,=\,2$ modes, green squares
$\ell\,=\,4$ and 9 modes, and red stars indicate periods which are not
associated with a mode. Periods which could be either $\ell\,=\,1$
or 2 are indicated with an extra colored circle (black in the left
panel and blue in the right panel). Point sizes are (logarithmically)
scaled with amplitude and vertical lines indicate the extent of
frequency multiplet splittings.}
\label{fig05}
\end{figure*}

Next we produced Echelle diagrams using the spacings indicated by the
KS test. These are shown in Fig.\,\ref{fig05} with different symbols
indicating mode identifications.
Even with these tools, we 
imposed additional conditions on our search for period sequences. 
We began by examining i) the high-amplitude periodicities
under the assumption that they are $\ell\,=\,1$ (as indicated by the 
strong peak in the KS test) and ii) using the multiplets
(both $\ell\,=\,1$ and 2) as starting points for sequences. 
We could easily identify a 
sequence for both $\ell\,=\,1$ and 2 modes.  We detected 35 $\ell\,=\,1$ 
periodicities and 95 $\ell\,=\,2$ periodicities that follow sequences within 
a small margin. We assigned radial indices ($n$) to 
the members of these sequences, estimating that the radial
fundamental mode would appear near 600\,seconds; a rough estimate for
stars with $\log g\approx 5.2$.  
This resulted in period spacings of $266.4\,\pm\,0.2$ and $153.2\,\pm\,0.2$
seconds for $\ell\,=\,1$ and 2, respectively, in good agreement with
the survey results of \citet{reed10a}.

Our best-estimate mode identifications, using multiplets and period spacings,
 are provided in Columns\,5 through 8 of Table\,\ref{tab01}.

\section{Discussion}

Now that mode identifications have been established, they can be 
examined using the tools applied in our previous papers.

\subsection{Unidentified Periodicities}

Ideally, every periodicity detected in KIC 3527751 would be associated
with a mode  or a combination frequency.  
With some degree of certainty, we have identified three
$\ell\,=\,4$ multiplets, and possibly even an $\ell\,=\,9$ multiplet, 
which invites a search for more high-degree
periodicities. To do this, we prewhiten the KS test. We remove our
assigned $\ell\,=\,1$ modes, $\ell\,=\,2$ modes, and then both. Of the
79 periodicities longer than 800\,s not identified as $\ell\,=\,1$ or 2, 
19 are identified
as $\ell\,=\,4$ and 13 as part of a possible $\ell\,=\,9$ multiplet, 
leaving just 45 of 225 $g$-mode periodicities ($f\,<\,1\,100\,\mu$Hz) 
as unidentified.
The bottom panels of Fig.\,\ref{fig04} show the period prewhitening
sequences. When the $\ell\,=\,2$ sequence is removed, a second peak
appears shortward of the $\ell\,=\,1$ peak, which could indicate
deviations or multiplet structure in that sequence. However, when both
$\ell\,=\,1$ and 2 are removed, some of that peak remains, indicating
that we may not be catching all of the $\ell\,=\,1$ sequence. However, no
new peaks appear where we would expect the higher-degree sequences
to be
(indicated by vertical lines near 109 and 84\,seconds for the $\ell\,=\,3$
and 4 sequences, respectively). The
three $\ell\,=\,4$ multiplets we detect have spacings between them of
257 ($3\times 86$) and 223 ($3\times 74$) seconds which are close to
$3\times$ the expected value. Yet we have to conclude that the
KS plots do not reveal obvious peaks where $\ell\,=\,3$
or 4 modes should be. Since many of the unidentified periods are
short (53 of these have periods below 1\,800\,s), they could well be  
$\ell\,=\,1$ and 2 
modes which do not fit the period sequences because they are low radial ($n$)
order.

\subsection{Trapped Modes}
\begin{figure*}[!htbp]
\figurenum{7}
\includegraphics[angle=-90,width=5.0in]{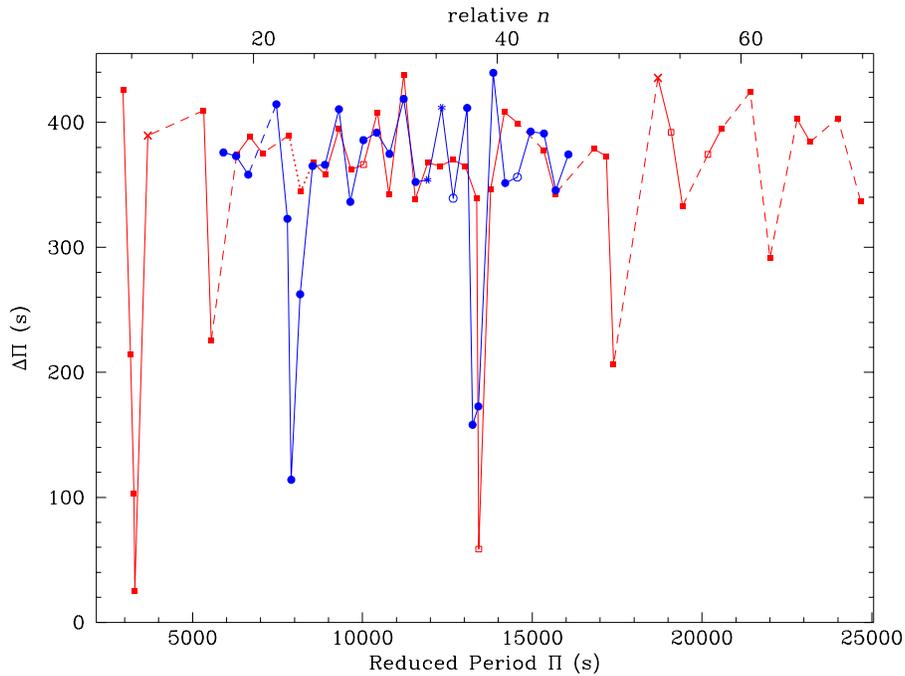}
\caption{Reduced periods, $\Pi\,=\,P\sqrt{\ell\left(\ell\,+\,1\right)}$,
indicating trapped modes. Circles (blue) indicate the $\ell\,=\,1$
sequence and squares (red) indicate the $\ell\,=\,2$ sequence.
Open symbols indicate periodicities discovered as the result of a search
for missing overtones and trapped $\ell\,=\,2$ modes, as discussed in the text,
and crosses indicate modes with the same period as that of a different
mode (e.g. an $\ell\,=\,1$ mode previously identified as $\ell\,=\,2$).
Dashed lines indicate missing overtones and dotted lines in the 
$\ell\,=\,2$ sequence indicate that a trapped mode is missing. }
\label{fig06}
\end{figure*}

Structural models have indicated significant mode trapping in the He--H
transition zone of sdB stars \citep{mibert,hu09,charp02a}. 
Period spacing sequences argued
against significant mode trapping \citep[see the discussion in \S 3 of
][]{reed11c} until the recent discovery of trapped modes in the sdBV star
KIC\,10553698A \citep{ostensen14a}.
Thus it is prudent to search for trapped modes in each
sdBV star analyzed. Unfortunately, not finding trapped modes
is like not finding a binary star- one can have positive
detections, but it is nearly impossible to conclusively rule them out.  For
KIC\,3527751, there are four periodicities ($f023$, $f024$, $f084$, 
and $f085$) with $\ell\,=\,1$-like
multiplet splittings which do not fit into the period sequence.
These can be seen in the Echelle diagram (Fig\,\ref{fig05}), but are
most easily seen in a reduced period plot (Figure\,\ref{fig06}), 
such as that in \citet{ostensen14a}. Reduced period plots are
also commonly shown in modeling papers, such as those previously
referenced. Periods are
converted to degree-independent reduced periods by multiplying by
$\sqrt{\ell\left(\ell\,+\,1\right)}$, which also makes the
period spacings degree independent.
The trapped $\ell\,=\,1$ modes
deviate shortward from the main spacing near 380\,seconds.

However, there are no corresponding trapped $\ell\,=\,2$ modes. In
regular period, the positions where the trapped
$\ell\,=\,2$ modes would occur are near 3270
and 5430\,seconds. There are no available periodicities near
3270\,seconds ($306\,\mu$Hz) even when looking below the $4\sigma$ limit.
So there are \emph{no} candidates for a
trapped $\ell\,=\,2$ mode corresponding
to this trapped $\ell\,=\,1$ mode.
For the second trapped mode,
there is a single peak at 5457\,seconds that we assign
as a trapped $\ell\,=\,2$ mode corresponding to the
trapped $\ell\,=\,1$ mode
near a reduced period of 13\,000, but mark it in
Fig.\,\ref{fig06} with an open symbol. 

Since the sequences were not
complete, we also looked for periodicities of missing overtones, to
complete the sequences. A few more possible periods were detected with 
amplitudes between 4 and $5\sigma$, and
are marked with open symbols, and we also detected some which overlap
the other sequence. Equation\,3 predicts that sequences overlap,
so mode degeneracy is expected. We have marked those periods
with crosses and note that these mode identifications are not favored because
of multiplet splittings. 

The $\ell\,=\,2$ sequence
extends well beyond the $\ell\,=\,1$ sequence and there are some 
indications of trapped modes at either end. Of course at the short end,
the radial orders are small, where asymptotic behavior is not expected
anyway and at the longer end, there are many missing radial orders, so
a lot of ambiguity exists. 
We include them for completeness. 

This would be the second
sdBV star to show strong indications of trapped modes, contrary to
what was indicated in \citet{reed11c}.

\subsection{Radially Differential Rotation}

\begin{figure*}[!htbp]
\figurenum{8}
\includegraphics[width=5.0 in]{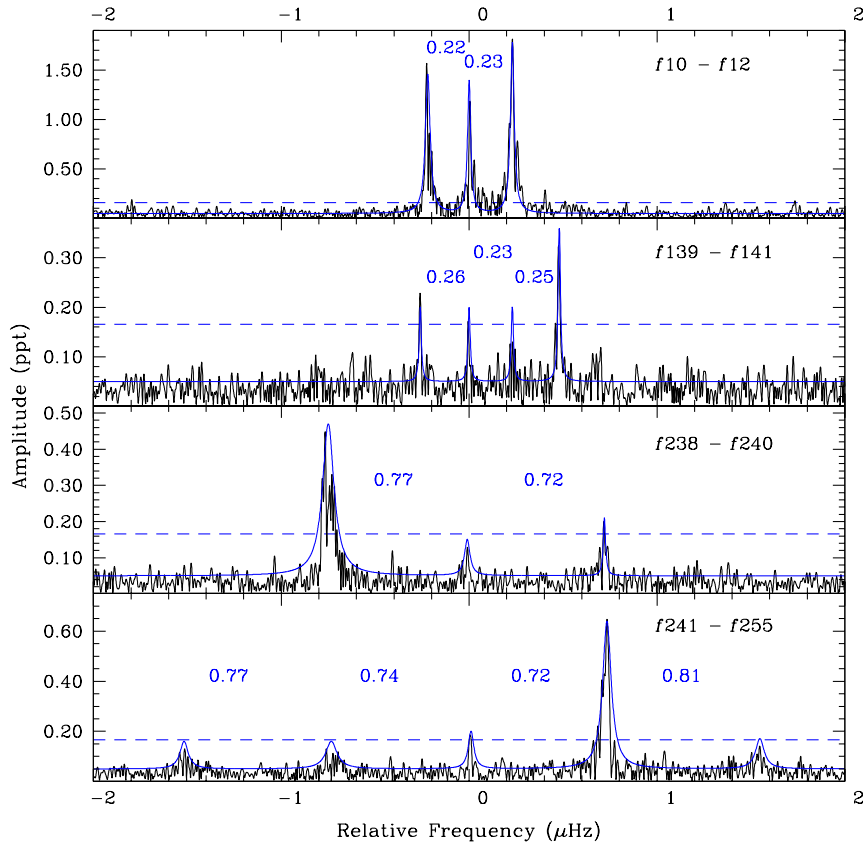}[t]
\caption{Frequency multiplets indicating radially differential rotation. The
top two panels show $g$-mode multiplets and the bottom two panels
$p$-mode multiplets. Numbers indicate the spacings between frequencies.}
\label{fig07}
\end{figure*}

Previous papers \citep[e.g.][]{baran12a} have used multiplets to determine
stellar rotation periods and since many multiplets were
detected in KIC\,3527751, we do it here too. We separate
the pulsations into $p$- and $g$-modes and further separate the
$g$-modes by degree, $\ell$. We do this because the splitting
in $g$-modes is degree-dependent whereas $p$-modes have small
Ledoux constants \citep[e.g.][]{vangrootel_F48}. For the $g$-modes,
we calculated the stellar rotation  period 
to be $46\,\pm\,8$, $41\,\pm\,5$,
and $40\,\pm\,5$\,days for the $\ell\,=\,1$, 2, and 4 multiplets,
respectively. These are all in close agreement.
For the $p$-modes, using a Ledoux constant of zero,
we calculated a stellar rotation period of $15.3\,\pm\,0.7$\,days.
We note that while several members of the two $p$-mode multiplets 
have low amplitudes, the frequency splittings are remarkably uniform
with  comparatively large splittings of $0.74\,\mu$Hz. These are shown in
the bottom two panels of Fig.\,\ref{fig07} with two $g$-mode
multiplets in the top panels.

\notetoeditor{color figure in online edition}

Here we arrive
at a contradiction; the two $p$-mode multiplets indicate a spin period of
$15.3\,\pm\,0.7$\,days while the $g$-mode multiplets indicate a period near
43\,days. We can imagine three possibilities which would generate these
results; i) the intrinsic $p$-mode multiplets are actually spaced at
$1/3$ our observed value, ii) KIC\,3527751 is
a pair of sdBV stars with one a $p$-mode pulsator rotating with a period near
15\,days and the other a $g$-mode pulsator rotating with a period near 
43\,days, or iii) KIC\,3527751 is differentially rotating. We discuss
all three possibilites below, but as to not leave the reader in too much
suspense, we discount options i and ii and highly favor option iii.

For option i to be correct, the $f239$ - $f241$ triplet would need to be
$\ell ,\,m\,=\,3,-3$; 3,0; and $3,+3$ and the $f242$ - $f246$ quintuplet
would need to be $\ell ,\,m\,=\,6,-6$; $6,-3$; 6,0; $6,+3$; and $6,+6$.
This seems quite unlikely since both of these have quite high geometric
cancellation factors \citep{pes85}.  In addition, there is no pulsation
inclination where the $m\,=\,1,\,2,\,4,$ and 5 modes have
a common node. Yet we cannot summarily rule this out
as we seem to detect $\ell\,=\,4$ and possibly 9 modes in the $g$-mode regime.
However, in the $g$-mode cases, there are many more higher-amplitude
periodicities whereas in the $p$-mode range there would only be one. As
such, we would have to question why the $\ell\,=\,0$, 1, and 2 modes are
not excited to observable levels when their cancellation factors are so
much smaller. Additionally, the frequency/amplitude pattern is similar
to what has been observed in the sdBV star Balloon\,090100001 (as discussed
in \S 3.1), making option i unlikely.

Option ii seems unlikely from the outset as an sdB+sdB
binary has never been observed, and only about 80\% of cool sdB stars pulsate
with $g$-modes and 10\% of hotter sdB stars pulsate with $p$-modes.
We could rule out this possibility immediately if members
of $p$- and $g$-mode pulsations were in combination frequencies. Unfortunately,
that is not the case as our possible combination frequencies are $g$-mode
(or mixed character) periodicities only (Table\,2).
Several other sdB stars have been discovered to be binary via
their \emph{Kepler} lightcurves which show signs of Doppler beaming,
tidal distortions, and/or eclipses \citep[e.g.][]{telting12a,bloemen12}.
These effects should only occur for relatively short-period binaries, and
so we re-processed KIC\,3527751's data prior to long-term trend removal
and this time detrended, month-by-month, only for periods longer than eight
or ten days. This detrending leaves 
three, closely-spaced, low-amplitude peaks in the FT
at 3.933, 3.962, and 4.026$\mu$Hz (2.94, 2.92, and 2.87\,days, respectively)
with amplitudes of 0.051, 0.042, and 0.056\,ppt, respectively. Already, with
three peaks, these are highly unlikely to be signatures of binarity, but
are most likely spurious noise which was not properly removed. But
to complete this exercise, using the Doppler beaming procedure of 
\citet[][specifically equation 2]{telting12a}, we calculate a 
 velocity amplitude of the binary's orbital motion
of 23.6\,km$\cdot$s$^{-1}$ from the flux amplitudes. 
From our 23 RV measurements, we determine
an RV amplitude of $16.5\,\pm\,3.2$\,km$\cdot$s$^{-1}$ for this
period. These velocities are not consistent with each other, and assuming
sdB stars of canonical mass ($0.48\,M_{\odot}$) would require inclinations
below $15^o$. This combination of evidence
makes it extremely unlikely that KIC\,3527751 is a binary pair
of pulsating sdB stars.

Therefore it is far more likely that KIC\,3527751 is differentially 
rotating  (option iii) and 
we consider this to be the case.
$p$-mode pulsations only sample the envelope whereas 
$g$-mode pulsations penetrate the He core and so are sensitive to conditions
deeper within the star
\citep[see example propogation diagrams in][]{charp14}.
As such, the $g$-mode-derived rotation rate is sensing deep into the 
star and the $p$-mode rate is sensing the outer regions. There is
overlap in the propagation diagram of \citet{charp14}, so the core
rate is likely slower than indicated by the $g$ modes, but these
results conclude that
the core is rotating at least three times slower than the envelope. To
determine a more precise value would require appropriate models constructed
using the observational constraints (mode IDs, period spacing/trapping
sequences, etc.) of this paper to calculate the proper propagation
diagram. To date, no detailed models have been published which fully 
incorporate the detailed seismic constraints provided from Kepler data.
Interestingly, a similar discovery has been detected in \emph{Kepler} 
observations of a pulsating main sequence A star 
\citep{kurtz14}, although in that case,
the core is only rotating about 5\% slower than the envelope. Radially
differential rotation has also been measured in red horizontal branch
stars, though in that case the core is rotating faster than the
envelope \citep{beck12}.

\subsection{Pulsation Inclination Angle}

As shown
in \citet{charpinet11b,reed05,pes85}, geometric cancellation is
inclination and $m$-dependent, which means we can determine the inclination
angle of the pulsation axis to our line-of-sight using multiplet amplitudes. 
For observations which span several amplitude
e-folding timescales, it is reasonable to presume that all amplitudes of all 
multiplet members have reached average amplitudes and
then their relative heights indicate the inclination angle. However,
as Fig.\,\ref{fig02} indicates, for
 KIC\,3527751 it does not appear that the data span more than a single amplitude
variation cycle, and so this assumption is likely not valid. Yet we can place
constraints using surface nodes, where geometric cancellation is complete, to
omit inclination ranges. The $\ell\,=\,1$ multiplet amplitudes, while
inconsistent between multiplets, have roughly equal amplitudes for all members,
 indicative of intermediate inclinations (30 -- 60$^o$). 
Several of the $\ell\,=\,2$
multiplets are quadruplets, a few are full quintuplets, and some are only triplets. Since
all members are present in various multiplets, it indicates inclinations above
$15^o$ (which would not have $m\,\neq\,0$ components)  and below $75^o$ (which would
suppress $m\,=\,2$ components). Additionally, there is an $m\,=\,0$ node line
at $i\,=\,54^o$ which eliminates inclinations between
50 and 65$^o$.
The $\ell\,=\,4$ multiplets are more sensitive to the inclination angle, yet are
inconsistent between multiplets. The $566\,\mu$Hz multiplet has only even $m$
members while the other two have roughly-equal amplitudes for all members. As such,
we have to presume that no members are suppressed. Since 
$\ell\,=\,4$ node lines appear at
$i\,=\,0$, 32, 48, 67, 70, and 90$^o$, inclinations within a
few degrees would be suppressed. 
What is left are inclinations between 35 and 45$^o$, based solely on multiplet
members being visible.

\subsection{Pulsation Density}
We can also examine where the pulsations seem to be most
easily driven. Near instability boundaries, it would be expected that
fewer periodicities would be driven and so mapping pulsation density
with overtone may provide constraints on the driving region itself
\citep[as in Fig.\,9 of][]{jeffery06a}.
We use multiplet members as a proxy of driving power; the
more multiplet members observed, the more driving power we presume
that mode to have. There are methods for doing this involving pulsation
amplitudes \citep[e.g.][]{mukadam2006}, however considering the variability
observed over the duration of the observations, this seems a difficult
proposition- Should we calculate the total power when the amplitudes
are maximum or median? Should we consider multiple times per multiplet
when each member is
maximum, or choose only one time when most members are visible? The scheme
we decided to use, number of multiplet members, is simpler in two ways:
it is binary- either multiplet
members are detected or they are not, and it imposes no time or amplitude
constraints. The members do not have to be simultaenously detected. 
Once structure
and pulsation models are sufficiently mature to accurately reflect the 
observed pulsation behavior, our scheme may 
prove not to be the best. However, with the
information currently in hand, it seems more robust than using 
amplitudes and a good beginning
point for determining pulsation instability.

\begin{figure}[!htbp]
\figurenum{9}
\plotone{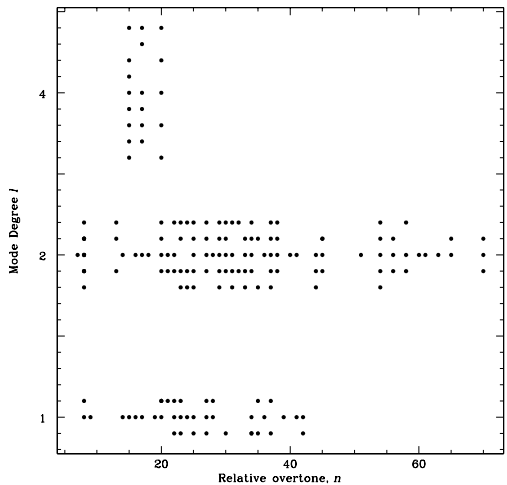}
\caption{Periodicities arranged by degree and radial index used to indicate
where maximum driving occurs.}
\label{fig08}
\end{figure}

Figure\,\ref{fig08} shows multiplet members by radial
order with the densest region between $n\,=\,20$ and 40 for $\ell\,=\,1$ and 2
while the $\ell\,=\,4$ multiplets have $15\leq n\leq20$. It is interesting
that the $\ell\,=\,4$ modes have lower orders as they should be very 
sensitive to pulsation power to drive them to observable amplitudes. Yet
as we are not producing models, we only indicate their relative regions 
as another constraint for those who do. Once all the \emph{Kepler}-observed
sdBV stars are analyzed, a comparison between different pulsators should
map the instability region.

\section{Summary and Conclusions}

We have detected 251 periodicities in KIC\,3527751 using 38 months of 
virtually continuous \emph{Kepler} data. Most of these are in the $g$-mode
regime of long periods, 14 should be $p$-modes with  periodicities shorter than 
six minutes, and 38 periods are between 9 and 20\,minutes and could
be modes of mixed character. We recovered all but one of the periodicities
detected by \citet{reed10a}, including those in the $p$-mode regime. We
do see two peaks near the 
missing frequency, at $544\,\mu$Hz in \citet{reed10a}, but since they are
both slightly below our $5\sigma$ limit, we did not list them.

Similar to most other \emph{Kepler}-observed sdBV stars, amplitudes and sometimes
frequency/phase variations occur throughout the extended dataset, making
traditional frequency-fitting and prewhitening tools unfeasible. We
therefore used sFTs and Lorentzian fitting to attempt to disentangle
the frequency content. Even so, there are regions which are extremely
messy, with amplitude variations and frequency variations which make
modes seemingly cross each other. For these regions, extractions are 
somewhat subjective as portions of FTs are chosen from the sFT for which
the pulsations are most clearly seen. This usually means the frequencies
are at their most separated which could apply a bias to the results. 
 We encourage others to examine the accompanying figures of sFTs,
or even better develop their own methods, for determining frequencies
and amplitudes, and their variations.

In an attempt to understand the pulsations, we applied a wide variety
of tools which have previously been used. Some of these did not reveal
new relationships: prewhitening the KS test did not show any
peaks indicative of $\ell\,\geq\,3$ modes and the density of mode multiplets
(Fig.\,\ref{fig08}) are not consistent between degrees. All three of the 
mode identification methods did work well, resulting
in 75\% of the periodicities having assigned modes. 

Our multiplet analysis also revealed three $\ell\,=\,4$ multiplets and
possibly an $\ell\,=\,9$ multiplet. 
Several other Kepler-observed sdB pulsators also have high-degree
($\ell\,\geq\,3$) modes, including KIC\,7668647 which has an $\ell\,=\,8$
multiplet \citep{jht14}. High-degree modes have 
long been a staple of sdB asteroseismology
to explain the density of periodicities of $p$-mode pulsations 
\citep[see the discussion of ][ \S 5.3 and Fig.\,15]{reed07a}. The precision
and duration of \emph{Kepler} data allowed us to detect periodicities with amplitudes
 $1/45$ of the highest one. With this large dynamic range,
we \emph{should} be seeing high-degree modes, yet only with confirming multiplet
structure can we be certain. As such, even Kepler-observed sdB pulsators
have a large number of unidentified periodicities which could be high-degree modes,
but there is no way to be certain. 

Since the Ledoux
constant is degree-dependent for $g$-modes, the rotation period ($1/\Omega$)
determined from different degrees can be used to check if the multiplet
splittings agree with predictions from Eqns.\,1 and 2. As stated in \S 4.3,
the rotation periods from the $\ell\,=\,1$, 2, and 4 modes are calculated
to be $46\,\pm\,8$, $41\,\pm\,5$, and $40\,\pm\,5$ days, respectively.
Since these periods all agree,
the mode-dependent Ledoux constants fit those predicted from Eqn\,2,
supporting our mode identifications.

Remarkable features in KIC\,3527751 include mode trapping and
radially differential rotation. The case for mode trapping is less
sure than that of \citet{ostensen14a}, where long complete sequences of
overtones were found.
In our case, there are several missing and
so we have had to average across those missing overtones. While
we have two nice $\ell\,=\,1$ trapped regions, with mode identifications
based on multiplet splittings, we do
not detect one of the corresponding $\ell\,=\,2$ trapped modes and the
other occurs where a non-trapped $\ell\,=\,1$ mode is, making its
identification uncertain. So we have good examples of $\ell\,=\,1$
trapped modes, but not for $\ell\,=\,2$.

We detect $g$-mode multiplet splittings for three degrees $\ell$ which
provide the same rotation period which is substantially different
from that derived from two $p$-mode multiplets. We examined the prospect
that KIC\,3527751 could actually be a binary set of pulsating stars,
and while we cannot completely rule it out, it is exceedingly unlikely,
as is the possibility that the $p$-mode multiplets are really spaced
like the $g$-mode ones, but are just missing members.
Therefore, we suggest that KIC\,3527751 is differentially rotating
in the radial direction with the core rotating nearly three times
more slowly. This 
is opposite to that described by \citet{kaw05}, who
predicted rapidly rotating cores via conservation of angular momentum
in a contracting core. This could be a clue to the mass loss mechanism
itself and a bit of a dichotomy. Either the core must preferentially
spin down during mass loss, or the entire star must spin down with the 
envelope subsequently being spun-up. The differential rotation of
KIC\,3527751 allows various mechanisms to be tested.

\acknowledgements

Funding for this research was provided by the National Science Foundation
grants \#1009436 and \#1312869. Any opinions, findings, and conclusions or
recommendations expressed in this material are those of the
author(s) and do not necessarily reflect the views of the National
Science Foundation. HF was
supported by the Missouri Space Grant Consortium,
funded by NASA.
This work was partially supported by Polish National Science Center
under project No.\,UMO-2011/03/D/ST9/01914.

This paper includes data collected by the \emph{Kepler} mission. Funding for the \emph{Kepler} mission is provided by the NASA Science Mission directorate.
Some/all of the data presented in this paper were obtained from the Mikulski Archive for Space Telescopes (MAST). STScI is operated by the Association of Universities for Research in Astronomy, Inc., under NASA contract NAS5-26555. Support for MAST for non-HST data is provided by the NASA Office of Space Science via grant NNX13AC07G and by other grants and contracts.

The spectroscopic observations used in this work were obtained with the
Mayall Telescope of Kitt Peak National Observatory, which is operated by the
Association of Universities for Research in Astronomy under cooperative
agreement with the National Science Foundation; the William Herschel
Telescope located at the Observatorio del Roque de los Muchachos (ORM)
and operated
by the Isaac Newton Group; and the Nordic Optical Telescope also
at the ORM and operated jointly by Denmark, Finland, Iceland, Norway, and
Sweden.

We thank the Referee for suggestions which improved the paper.

\bibliography{sdbrefs}

\clearpage
\appendix
\setcounter{table}{0}
\begin{deluxetable}{lcccccccccc}
\tablecaption{Table of asteroseismic quantities. Column\,1 provides a 
label for the periodicity, Columns\,2 and 3 the frequency and period,
with errors in parentheses, and column\,4 the observed amplitude. Columns\,5
through 8 provide our best estimate mode identifications, Columns\,9
and 10 period spacing deviations, and Column\,11 the frequency splitting
(from the subsequent frequency) of multiplet members. $^{\dag}$ indicates
periodicities detected by \citet{reed10a} and $^*$ indicates frequencies which
are listed in Table\,\ref{combo} as part of a combination frequency. 
Periodicities listed with ``t''s instead of ``f''s in the
ID column are below the $5\sigma$ limit
and therefore considered \emph{tentative}.}
\tablehead{
\colhead{ID} & \colhead{Freq.} & \colhead{Period} & \colhead{Amp.} &
\colhead{$\ell$} & \colhead{$m$} & \colhead{$n_{\ell\,=\,1}$} & \colhead{$n_{\ell\,=\,2}$} &
\colhead{$\left(\frac{\Delta P}{P}\right)_{\ell\,=\,1}$} & \colhead{$\left(\frac{\Delta P}{P}\right)_{\ell\,=\,2}$} & \colhead{$\delta f$}\\
& \colhead{$\mu$Hz} & \colhead{sec.} & \colhead{ppt} &\nodata &\nodata &\nodata &\nodata &\nodata &\nodata & \colhead{$\mu$Hz} }
\startdata
f001 & 72.026 (0.015) & 13883.84 (2.91) & 0.17 &  &  &  &  &  &  &  \\
f002$^*$ & 76.174 (0.010) & 13127.84 (1.68) & 0.16 &  &  &  &  &  &  &  \\
f003 & 77.630 (0.024) & 12881.62 (3.99) & 0.40 &  &  &  &  &  &  &   \\
f004 & 84.314 (0.006) & 11860.43 (0.86) & 0.16 &  &  &  &  &  &  &  \\
f005 & 84.790 (0.006) & 11793.84 (0.81) & 0.17 &  &  &  &  &  &  &  \\
f006 & 85.428 (0.016) & 11705.79 (2.16) & 0.24 &  &  &  &  &  &  &   \\
f007 & 87.851 (0.008) & 11382.91 (1.04) & 0.24 & 1 & -1 & 40 &  & 0.06 &  & 0.1809 \\
f008 & 88.032 (0.010) & 11359.52 (1.31) & 0.32 & 1 & 0 & 40 &  & -0.02 &  &   \\
f009 & 90.130 (0.019) & 11095.08 (2.34) & 0.20 & 1 &  & 39 &  & -0.02 &  &   \\
f010 & 91.703 (0.008) & 10904.73 (0.91) & 1.64 & 2 & -1 &  & 68 &  & 0.35 & 0.231 \\
f011 & 91.934 (0.007) & 10877.33 (0.85) & 1.24 & 2 & 0 &  & 68 &  & 0.18 & 0.2266 \\
f012$^{\dag}$ & 92.161 (0.010) & 10850.59 (1.21) & 1.89 & 2 & 1 & 38 & 68 & 0.07 & 0.00 &   \\
f013 & 94.590 (0.032) & 10571.94 (3.56) & 0.38 & 1 &  & 37 &  & 0.02 &  &   \\
f014 & 99.289 (0.004) & 10071.59 (0.41) & 0.22 & 2 & 0 &  & 63 &  & -0.08 & 0.1221 \\
f015$^*$ & 99.411 (0.005) & 10059.22 (0.47) & 0.19 & 1 & 1 & 35 & 63 & 0.09 & -0.16 & 0.1141 \\
f016 & 99.525 (0.007) & 10047.69 (0.69) & 0.23 & 2 & 1 &  & 63 &  & -0.24 & 0.176 \\
f017 & 99.701 (0.012) & 10029.95 (1.18) & 0.22 & 1 & -1 & 35 & 63 & -0.02 & -0.35 &   \\
f018 & 101.589 (0.020) & 9843.59 (1.94) & 0.26 &  &  &  &  &  &  &   \\
f019 & 102.083 (0.006) & 9795.95 (0.61) & 0.21 &  1 or 2 &  & 34 & 61 & 0.11 & 0.12 &   \\
f020 & 105.304 (0.014) & 9496.32 (1.22) & 0.63 & 1 & -1 & 33 &  & -0.02 &  & 0.248 \\
f021$^{\dag}$ & 105.552 (0.019) & 9474.00 (1.68) & 1.71 & 1 & 1 & 33 &  & -0.10 &  & 0.08 \\
f022$^*$ & 105.632 (0.012) & 9466.83 (1.08) & 2.92 & 2 & 0 &  & 59 &  & -0.03 &   \\
f023 & 106.678 (0.008) & 9374.00 (0.70) & 0.20 & 1 & -1 &  &  &   &  & 0.128 \\
f024 & 106.806 (0.019) & 9362.77 (1.67) & 0.23 & 1 & 0 &  &  &   &  &   \\
f025 & 107.415 (0.015) & 9309.69 (1.30) & 0.39 & 2 & 0 &  & 58 &  & -0.05 &   \\
f026 & 108.085 (0.010) & 9251.98 (0.86) & 0.29 & 1 & -1 & 32 &  & 0.06 &  &   \\
f027 & 111.155 (0.036) & 8996.45 (2.93) & 0.32 & 2 & -1 &  & 56 &  & -0.10 & 0.182 \\
f028 & 111.337 (0.030) & 8981.74 (2.45) & 0.35 & 2 & 0 & 31 & 56 & 0.05 & -0.19 & 0.566 \\
f029 & 111.903 (0.005) & 8936.31 (0.42) & 0.48 & 2 & 2 &  & 56 &  & -0.49 &   \\
f030 & 114.062 (0.009) & 8767.16 (0.68) & 1.00 & 2 & -1 &  & 54 &  & 0.41 & 0.307 \\
f031 & 114.369 (0.006) & 8743.63 (0.49) & 0.92 & 2 & 0 &  & 54 &  & 0.25 & 0.291 \\
f032$^{\dag}$ & 114.660 (0.014) & 8721.44 (1.05) & 1.41 & 2 & 1 &  & 54 &  & 0.11 &   \\
f033 & 116.204 (0.004) & 8605.56 (0.31) & 0.19 &  &  &  &  &  &  &   \\
f034 & 118.624 (0.008) & 8430.00 (0.56) & 0.27 & 2 & -2 &  & 52 &  & 0.21 & 0.242 \\
f035$^{\dag}$$^*$ & 118.866 (0.012) & 8412.83 (0.82) & 1.31 & 2 & -1 &  & 52 &  & 0.09 & 0.218 \\
f036 & 119.084 (0.010) & 8397.43 (0.71) & 0.43 & 2 & - &  & 52 &  & -0.01 & 0.278 \\
f037 & 119.362 (0.013) & 8377.88 (0.91) & 0.26 & 2 & 1 &  & 52 &  & -0.14 & 0.229 \\
f038 & 119.591 (0.007) & 8361.83 (0.49) & 0.23 & 2 & 2 &  & 52 &  & -0.24 &  \\
f039 & 122.260 (0.120) & 8179.29 (8.02) & 0.22 & 1 & -1 & 28 &  & 0.04 &  &  \\
f040$^{\dag}$ & 126.095 (0.023) & 7930.53 (1.45) & 1.15 & 2 & 0 &  & 49 &  & -0.05 &  \\
f041 & 130.993 (0.007) & 7634.00 (0.41) & 0.19 & 1 & 0 & 26 & 47 & -0.01 & 0.01 & 0.116 \\
f042 & 131.109 (0.015) & 7627.24 (0.87) & 0.18 & 1 & 1 & 26 & 47 & -0.03 & -0.03 &   \\
f043 & 135.606 (0.013) & 7374.30 (0.71) & 0.35 & 1 & -1 & 25 &  & 0.02 &  & 0.104 \\
f044$^{\dag}$$^*$ & 135.710 (0.017) & 7368.65 (0.91) & 0.35 & 1 & 0 & 25 &  & -0.01 &  & 0.112 \\
f045 & 135.822 (0.015) & 7362.58 (0.81) & 0.48 & 1 & 1 & 25 &  & -0.03 &  &   \\
f046 & 140.530 (0.045) & 7115.92 (2.28) & 1.28 & 2 & -1 &  & 43 &  & 0.63 & 0.3 \\
f047 & 140.830 (0.045) & 7100.76 (2.27) & 1.22 & 2 & 0 &  & 43 &  & 0.53 & 0.163 \\
f048$^{\dag}$ & 140.993 (0.012) & 7092.55 (0.63) & 2.29 & 2 & 1 & 24 & 43 & -0.04 & 0.48 &   \\
f049 & 142.521 (0.007) & 7016.51 (0.33) & 0.30 & 2 & 1 &  & 43 &  & -0.02 &   \\
f050 & 144.362 (0.014) & 6927.03 (0.66) & 0.21 &  &  &  &  &  &  & 0.752 \\
f051 & 145.114 (0.005) & 6891.13 (0.24) & 0.20 & 2 & -2 &  & 42 &  & 0.16 & 0.299 \\
f052 & 145.413 (0.007) & 6876.96 (0.33) & 0.26 & 2 & -1 &  & 42 &  & 0.07 & 0.283 \\
f053 & 145.696 (0.024) & 6863.61 (1.14) & 0.45 & 2 & 0 &  & 42 &  & -0.02 & 0.065 \\
f054 & 145.761 (0.014) & 6860.55 (0.67) & 0.45 & 2 &  &  & 42 &  & -0.04 &  \\
f055 & 146.344 (0.012) & 6833.21 (0.58) & 0.29 &  &  &  &  &  &  & 0.186 \\
f056 & 146.530 (0.015) & 6824.54 (0.70) & 0.40 & 1 & -1 & 23 &  & -0.05 &  & 0.107 \\
f057 & 146.637 (0.009) & 6819.56 (0.40) & 0.43 & 1 & 0 & 23 &  & -0.07 &  &   \\
f058 & 148.156 (0.009) & 6749.64 (0.41) & 0.19 &  &  &  &  &  &  &   \\
f059 & 151.942 (0.015) & 6581.46 (0.66) & 0.38 & 1 & 0 & 22 &  & 0.04 &  & 1.739 \\
f060 & 153.681 (0.021) & 6506.99 (0.90) & 0.20 &  &  &  &  &  &  &   \\
f061 & 156.244 (0.009) & 6400.25 (0.37) & 0.27 & 2 & 0 &  & 39 &  & -0.04 &   \\
f062 & 157.706 (0.009) & 6340.91 (0.36) & 0.18 &  &  &  &  &  &  &   \\
f063$^*$ & 158.845 (0.036) & 6295.45 (1.44) & 1.51 & 1 & -1 & 21 &  & -0.03 &  & 0.107 \\
f064 & 158.952 (0.009) & 6291.21 (0.36) & 0.67 & 1 & 0 & 21 &  & -0.05 &  & 0.134 \\
f065$^{\dag}$ & 159.086 (0.011) & 6285.91 (0.43) & 0.71 & 1 & 1 & 21 &  & -0.07 &  &   \\
f066$^{\dag}$ & 159.731 (0.009) & 6260.53 (0.35) & 0.23 & 2 &  &  & 38 &  & 0.05 &   \\
f067 & 165.633 (0.007) & 6037.44 (0.26) & 0.42 & 1 & -1 & 20 &  & 0.00 &  & 0.142 \\
f068 & 165.775 (0.025) & 6032.27 (0.91) & 0.25 & 1 & 0 & 20 &  & -0.02 &  & 0.139 \\
f069 & 165.914 (0.006) & 6027.22 (0.20) & 0.35 & 1 & 1 & 20 &  & -0.04 &  &   \\
f070$^{\dag}$ & 167.786 (0.006) & 5959.97 (0.21) & 0.26 & 2 & -1 &  & 36 &  & 0.09 & 0.229 \\
f071 & 168.015 (0.017) & 5951.85 (0.60) & 0.26 & 2 & 0 &  & 36 &  & 0.03 & 0.202 \\
f072 & 168.217 (0.007) & 5944.70 (0.25) & 0.23 & 2 & 1 &  & 36 &  & -0.01 & 0.273 \\
f073 & 168.490 (0.035) & 5935.07 (1.23) & 0.17 & 2 & 2 &  & 36 &  & -0.08 &   \\
f074$^{\dag}$ & 171.908 (0.013) & 5817.06 (0.43) & 0.89 &  &  &  &  &  &  & 0.232 \\
f075 & 172.140 (0.030) & 5809.23 (1.01) & 0.70 &  &  &  &  &  &  & 0.189 \\
f076 & 172.329 (0.017) & 5802.85 (0.56) & 0.49 & 2 & -1 &  & 35 &  & 0.06 & 0.2 \\
f077 & 172.529 (0.011) & 5796.13 (0.39) & 0.68 & 2 & 0 &  & 35 &  & 0.02 & 0.211 \\
f078 & 172.740 (0.039) & 5789.05 (1.31) & 0.25 & 2 & 1 &  & 35 &  & -0.03 & 0.28 \\
f079 & 173.020 (0.019) & 5779.68 (0.63) & 0.41 & 2 & 2 &  & 35 &  & -0.09 & 0.161 \\
f080 & 173.181 (0.005) & 5774.31 (0.18) & 0.41 & 1 & 0 & 19 &  & 0.01 &  & 0.138 \\
f081 & 173.319 (0.016) & 5769.71 (0.52) & 1.39 & 2 & -2 &  & 35 &  & -0.16 &   \\
f082 & 177.871 (0.025) & 5622.05 (0.78) & 0.24 & 2 &  &  & 34 &  & -0.12 &   \\
f083 & 178.930 (0.009) & 5588.78 (0.28) & 0.19 & 1 & 0 & 18 &  & 0.31 &  & 0.088 \\
f084 & 179.018 (0.004) & 5586.03 (0.14) & 0.20 & 1 & 1 & 18 &  & 0.30 &  &   \\
f085 & 180.763 (0.012) & 5532.11 (0.37) & 0.45 &  &  &  &  &  &  & 0.261 \\
f086$^{\dag}$ & 181.024 (0.022) & 5524.13 (0.68) & 0.96 &  &  &  &  &  &  & 0.528 \\
f087 & 181.552 (0.006) & 5508.06 (0.18) & 0.20 & 1 & 1 & 18 &  & 0.01 &  & 0.508 \\
f088 & 182.060 (0.007) & 5492.69 (0.21) & 0.24 & 2 & -2 &  & 33 &  & 0.04 & 0.54 \\
f089 & 182.600 (0.017) & 5476.45 (0.51) & 0.25 & 2 & 1 &  & 33 &  & -0.07 & 0.652 \\
f090 & 183.252 (0.021) & 5456.97 (0.62) & 0.47 &  &  &  &  &  &  &   \\
f091 & 187.796 (0.023) & 5324.93 (0.65) & 0.45 & 2 & -1 &  & 32 &  & -0.06 & 0.246 \\
f092 & 188.042 (0.010) & 5317.96 (0.29) & 0.29 & 2 & 0 &  & 32 &  & -0.10 & 0.203 \\
f093 & 188.245 (0.010) & 5312.23 (0.29) & 0.19 & 2 & 1 &  & 32 &  & -0.14 & 0.233 \\
f094 & 188.478 (0.004) & 5305.66 (0.12) & 0.15 & 2 & 2 &  & 32 &  & -0.18 &   \\
f095 & 189.570 (0.006) & 5275.10 (0.17) & 1.53 & 1 &  & 17 &  & 0.14 &  &   \\
f096 & 193.031 (0.010) & 5180.52 (0.27) & 0.20 & 2 & -2 &  & 31 &  & 0.00 & 0.202 \\
f097 & 193.233 (0.023) & 5175.10 (0.62) & 0.20 & 2 & -1 &  & 31 &  & -0.04 & 0.205 \\
f098 & 193.438 (0.008) & 5169.62 (0.22) & 0.21 & 2 & 0 &  & 31 &  & -0.07 & 0.228 \\
f099$^*$ & 193.666 (0.007) & 5163.53 (0.19) & 0.22 & 2 & 1 &  & 31 &  & -0.11 &   \\
f100 & 198.960 (0.005) & 5026.14 (0.13) & 0.17 & 2 & -1 &  & 30 &  & -0.01 & 0.71 \\
f101 & 199.670 (0.005) & 5008.26 (0.13) & 0.18 & 2 & 2 &  & 30 &  & -0.12 &   \\
f102 & 204.698 (0.006) & 4885.25 (0.14) & 0.18 & 2 & -2 &  & 29 &  & 0.07 & 0.343 \\
f103$^{\dag}$ & 205.041 (0.007) & 4877.07 (0.18) & 0.22 & 2 & -1 &  & 29 &  & 0.02 & 0.332 \\
f104 & 205.373 (0.033) & 4869.19 (0.77) & 0.25 & 2 & 0 &  & 29 &  & -0.03 & 0.437 \\
f105 & 205.810 (0.006) & 4858.85 (0.14) & 0.20 & 2 & 2 &  & 29 &  & -0.10 & 0.739 \\
f106 & 206.549 (0.005) & 4841.47 (0.13) & 0.23 &  &  &  &  &  &  &   \\
f107 & 211.620 (0.007) & 4725.45 (0.17) & 0.43 & 2 & -1 &  & 28 &  & 0.03 & 0.264 \\
f108$^{\dag}$$^*$ & 211.884 (0.006) & 4719.56 (0.14) & 3.30 & 2 & 0 &  & 28 &  & -0.01 & 0.321 \\
f109 & 212.205 (0.007) & 4712.42 (0.15) & 0.52 & 2 & 1 &  & 28 &   & -0.06 & 0.282 \\
f110$^{\dag}$ & 212.487 (0.010) & 4706.17 (0.22) & 0.48 & 2 & 2 &  & 28 &   & -0.10 & 0.562 \\
f111 & 213.049 (0.017) & 4693.76 (0.37) & 0.70 & 1 &  & 15 &  & -0.05 &  &   \\
f112$^{\dag}$ & 217.834 (0.011) & 4590.65 (0.22) & 0.47 & 2 & -2 &  & 27 &  & 0.15 & 0.256 \\
f113 & 218.090 (0.015) & 4585.26 (0.32) & 0.87 & 2 & -1 &  & 27 &  & 0.11 & 0.187 \\
f114$^*$ & 218.277 (0.021) & 4581.33 (0.44) & 1.14 & 2 & 0 &  & 27 &  & 0.09 & 0.223 \\
f115$^{\dag}$ & 218.500 (0.037) & 4576.66 (0.77) & 0.78 & 2 & 1 &  & 27 &  & 0.06 & 0.253 \\
f116 & 218.753 (0.007) & 4571.37 (0.15) & 0.45 & 2 & 2 &  & 27 &  & 0.02 &   \\
f117 & 221.981 (0.032) & 4504.89 (0.65) & 0.21 &  &  &  &  &  &  & 3.221 \\
f118 & 225.202 (0.005) & 4440.46 (0.10) & 0.58 & 1 & 0 & 14 &  & 0.00 &  &   \\
f119 & 227.151 (0.011) & 4402.36 (0.21) & 0.36 & 2 &  &  & 26 &  & -0.08 &   \\
f120 & 234.423 (0.007) & 4265.79 (0.13) & 0.42 & 2 & -1 &  & 25 &  & 0.03 & 0.176 \\
f121$^{\dag}$ & 234.599 (0.010) & 4262.59 (0.19) & 0.44 & 2 & 0 &  & 25 &  & 0.01 & 0.209 \\
f122 & 234.808 (0.015) & 4258.80 (0.27) & 0.30 & 2 & 1 &  & 25 &  & -0.02 & 0.195 \\
f123 & 235.003 (0.006) & 4255.26 (0.10) & 0.28 & 2 & 2 &  & 25 &  & -0.04 &   \\
f124 & 239.422 (0.005) & 4176.73 (0.08) & 0.27 & 1 & 0 & 13 &  & 0.01 &  &  \\
f125 & 252.987 (0.025) & 3952.77 (0.39) & 1.79 & 2 & -2 &  & 23 &  & -0.01 & 0.207 \\
f126$^{\dag}$$^*$ & 253.194 (0.023) & 3949.54 (0.36) & 1.21 & 2 & -1 &  & 23 &  & -0.03 & 0.188 \\
f127 & 253.382 (0.012) & 3946.61 (0.19) & 0.53 & 2 & 0 &  & 23 &  & -0.05 & 0.199 \\
f128 & 253.581 (0.015) & 3943.51 (0.24) & 0.28 & 2 & 1 &  & 23 &  & -0.07 & 0.216 \\
f129 & 253.797 (0.008) & 3940.16 (0.12) & 0.20 & 2 & 2 &  & 23 &  & -0.09 &   \\
f130$^{\dag}$ & 255.686 (0.006) & 3911.05 (0.09) & 7.11 & 1 & 0 & 12 &  & 0.01 &  &   \\
f131 & 262.151 (0.008) & 3814.60 (0.11) & 0.50 &  &  &  &  &  &  & 0.099 \\
f132 & 262.250 (0.004) & 3813.16 (0.06) & 0.36 & 2 & -2 &  & 22 &  & 0.08 & 0.215 \\
f133$^{\dag}$ & 262.465 (0.008) & 3810.03 (0.12) & 0.35 & 2 & -1 &  & 22 &  & 0.06 & 0.779 \\
f134$^{\dag}$ & 263.244 (0.010) & 3798.76 (0.15) & 0.59 & 2 & 2 &  & 22 &  & -0.02 &  \\
f135 & 274.455 (0.016) & 3643.58 (0.21) & 0.18 & 2 & -2 & 11 & 21 & 0.01 & -0.03 & 0.245 \\
f136 & 274.700 (0.009) & 3640.33 (0.12) & 0.28 & 2 & -1 & 11 & 21 & 0.00 & -0.05 & 0.433 \\
f137$^{\dag}$ & 275.133 (0.021) & 3634.61 (0.27) & 0.59 & 2 & 1 &  & 21 &  & -0.09 & 0.223 \\
f138 & 275.356 (0.009) & 3631.66 (0.12) & 0.44 & 2 & 2 &  & 21 &  & -0.11 &   \\
f139 & 286.171 (0.005) & 3494.41 (0.06) & 0.23 & 2 & -1 &  & 20 &  & 0.00 & 0.255 \\
f140 & 286.426 (0.007) & 3491.30 (0.09) & 0.17 & 2 & 0 &  & 20 &  & -0.02 & 0.481 \\
f141 & 286.907 (0.007) & 3485.45 (0.09) & 0.32 & 2 & 2 &  & 20 &  & -0.06 &   \\
f142 & 298.961 (0.004) & 3344.92 (0.05) & 0.23 & 2 & -1 &  & 19 &  & 0.02 & 0.341 \\
f143$^*$ & 299.302 (0.005) & 3341.11 (0.05) & 0.42 & 2 & 0 &  & 19 &  & 0.00 &   \\
f144 & 312.217 (0.010) & 3202.90 (0.10) & 0.19 & 2 & -1 &  & 18 &  & 0.09 & 0.241 \\
f145 & 312.458 (0.014) & 3200.43 (0.14) & 0.16 & 2 & 0 &  & 18 &  & 0.08 & 0.241 \\
f146$^*$ & 312.699 (0.007) & 3197.96 (0.07) & 0.13 & 2 & 2 &  & 18 &  & 0.06 & 0.245 \\
f147 & 312.944 (0.007) & 3195.46 (0.07) & 0.16 & 2 & 1 &  & 18 &  & 0.05 &   \\
f148 & 346.949 (0.015) & 2882.27 (0.12) & 0.19 & 2 &  &  & 16 &  & 0.00 &   \\
f149 & 358.701 (0.005) & 2787.84 (0.04) & 0.25 &  &  &  &  &  &  &   \\
f150$^{\dag}$ & 366.419 (0.005) & 2729.12 (0.04) & 1.05 & 2 &  &  & 15 &  & 0.00 &   \\
f151$^{\dag}$ & 389.015 (0.005) & 2570.59 (0.03) & 0.68 &  1 or 2 &  & 7 & 14 & -0.02 & -0.03 &   \\
f152 & 427.770 (0.012) & 2337.70 (0.07) & 0.16 &  &  &  &  &  &  & 0.257 \\
f153 & 428.027 (0.013) & 2336.30 (0.07) & 0.20 & 1 & 0 & 6 &  & 0.10 &  &   \\
f154$^{\dag}$ & 441.507 (0.005) & 2264.97 (0.03) & 0.79 & 2 &  &  & 12 &  & -0.03 &   \\
f155 & 459.762 (0.042) & 2175.04 (0.20) & 0.17 & 2 & 2 &  & 11 &  & 0.39 & 0.234 \\
f156 & 459.996 (0.009) & 2173.93 (0.04) & 0.38 & 2 & -1 &  & 11 &  & 0.38 & 0.465 \\
f157 & 460.461 (0.007) & 2171.74 (0.03) & 0.21 & 2 & 1 &  & 11 &  & 0.36 &   \\
f158$^*$ & 565.893 (0.016) & 1767.12 (0.05) & 0.21 & 4 & -4 &  &  &  &  & 0.501 \\
f159$^{\dag}$ & 566.394 (0.006) & 1765.56 (0.02) & 1.21 & 4 & -2 &  &  &  &  & 0.526 \\
f160 & 566.920 (0.010) & 1763.92 (0.03) & 0.44 & 4 & 0 &  &  &  &  & 0.538 \\
f161 & 567.458 (0.014) & 1762.24 (0.04) & 0.25 & 4 & 2 &  &  &  &  & 0.319 \\
f162 & 567.777 (0.010) & 1761.25 (0.03) & 0.25 & 4 & 4 &  &  &  &  &   \\
f163 & 614.605 (0.009) & 1627.06 (0.02) & 0.21 &  &  &  &  &  &  &   \\
f164 & 662.979 (0.028) & 1508.34 (0.06) & 0.16 & 4 & -3 &  &  &  &  & 0.251 \\
f165 & 663.230 (0.014) & 1507.77 (0.03) & 0.19 & 4 & -2 &  &  &  &  & 0.219 \\
f166 & 663.449 (0.007) & 1507.27 (0.02) & 0.16 & 4 & -1 &  &  &  &  & 0.213 \\
f167$^{\dag}$ & 663.662 (0.029) & 1506.79 (0.07) & 0.20 & 4 & 0 &  &  & & & 0.778 \\
f168 & 664.440 (0.021) & 1505.03 (0.05) & 0.17 & 4 & 3 &  &  &  &  & 0.36 \\
f169 & 664.800 (0.015) & 1504.21 (0.03) & 0.16 & 4 & 4 &  &  &  &  &   \\
f170$^{\dag}$ & 719.582 (0.020) & 1389.70 (0.04) & 0.20 &  &  &  &  &  &  & 0.311 \\
f171 & 719.893 (0.008) & 1389.10 (0.02) & 0.18 &  &  &  &  &  &  &   \\
f172 & 743.009 (0.010) & 1345.88 (0.02) & 0.22 & 2 & -1 &  & 6 &  & -0.02 & 0.332 \\
f173 & 743.341 (0.012) & 1345.28 (0.02) & 0.25 & 2 & 0 &  & 6 &  & -0.03 & 0.232 \\
f174$^{\dag}$ & 743.573 (0.007) & 1344.86 (0.01) & 0.25 & 2 & 1 &  & 6 &  & -0.03 &   \\
f175$^{\dag}$ & 744.545 (0.010) & 1343.10 (0.02) & 0.23 &  &  &  & 6 &  & -0.04 &  \\
f176 & 748.392 (0.008) & 1336.20 (0.01) & 0.45 & 2 & -2 &  & 6 &  & -0.09 & 0.307 \\
f177 & 748.699 (0.009) & 1335.65 (0.02) & 1.19 & 2 & -1 &  & 6 &  & -0.09 & 0.315 \\
f178 & 749.014 (0.012) & 1335.09 (0.02) & 0.56 & 2 & 0 &  & 6 &  & -0.10 & 0.32 \\
f179$^{\dag}$ & 749.334 (0.017) & 1334.52 (0.03) & 0.75 & 2 & 1 &  & 6 &  & -0.10 & 0.38 \\
f180 & 749.714 (0.038) & 1333.84 (0.07) & 0.75 & 2 & 2 &  & 6 &  & -0.10 &   \\
f181 & 773.069 (0.010) & 1293.55 (0.02) & 0.32 & 2 & -1 &  & 6 &  & -0.37 & 0.366 \\
f182 & 773.435 (0.006) & 1292.93 (0.01) & 0.43 & 2 & 0 &  & 6 &  & -0.37 & 0.385 \\
f183 & 773.820 (0.011) & 1292.29 (0.02) & 0.50 & 2 & 1 &  & 6 &  & -0.37 & 0.329 \\
f184$^*$ & 774.149 (0.013) & 1291.74 (0.02) & 0.22 &  &  &  &  &  &  & 0.313 \\
f185 & 774.462 (0.033) & 1291.22 (0.06) & 0.28 &  &  &  &  &  &  & 0.214 \\
f186$^{\dag}$ & 774.676 (0.024) & 1290.86 (0.04) & 0.41 &  &  &  &  &  &  & 0.12 \\
f187 & 774.796 (0.016) & 1290.66 (0.03) & 0.43 &  &  &  &  &  &  & 0.452 \\
f188 & 775.248 (0.009) & 1289.91 (0.01) & 0.31 &  &  &  &  &  &  &   \\
f189 & 777.390 (0.040) & 1286.36 (0.07) & 0.50 &  &  &  &  &  &  & 0.623 \\
f190 & 778.013 (0.008) & 1285.326 (0.013) & 0.17 & 4 & -4 &  &  &  &  & 0.282 \\
f191 & 778.295 (0.008) & 1284.860 (0.013) & 0.15 & 4 & -3 &  &  &  &  & 0.299 \\
f192 & 778.594 (0.030) & 1284.366 (0.049) & 0.14 & 4 & -2 &  &  &  &  & 0.261 \\
f193 & 778.855 (0.005) & 1283.936 (0.009) & 0.25 & 4 & -1 &  &  &  &  & 0.283 \\
f194 & 779.138 (0.007) & 1283.470 (0.011) & 0.24 & 4 & 0 &  &  &  &  & 0.285 \\
f195 & 779.423 (0.008) & 1283.000 (0.014) & 0.25 & 4 & 1 &  &  &  &  & 0.297 \\
f196 & 779.720 (0.025) & 1282.512 (0.041) & 0.25 & 4 & 2 &  &  &  &  & 0.55 \\
f197 & 780.270 (0.007) & 1281.608 (0.011) & 0.12 & 4 & 4 &  &  &  &  &   \\
f198 & 823.716 (0.011) & 1214.011 (0.016) & 0.20 &  &  &  &  &  &  & 5.933 \\
f199 & 829.649 (0.006) & 1205.329 (0.008) & 0.20 & 2 &  &  & 5 &  & 0.06 &   \\
f200 & 887.122 (0.008) & 1127.241 (0.010) & 0.17 &  &  &  &  &  &  &   \\
f201 & 892.890 (0.007) & 1119.959 (0.009) & 0.23 &  &  &  &  &  &  & 0.612 \\
f202 & 893.502 (0.008) & 1119.192 (0.010) & 0.18 &  &  &  &  &  &  & 0.291 \\
f203$^{\dag}$ & 893.793 (0.030) & 1118.827 (0.038) & 0.19 &  &  &  &  &  &  &   \\
f204 & 931.652 (0.019) & 1073.362 (0.022) & 0.20 &  &  &  &  &  &  & 0.327 \\
f205 & 931.979 (0.006) & 1072.986 (0.006) & 0.20 &  &  &  &  &  &  & 0.365 \\
f206 & 932.344 (0.037) & 1072.565 (0.043) & 0.43 &  &  &  &  &  &  & 0.23 \\
f207$^{\dag}$ & 932.574 (0.020) & 1072.301 (0.023) & 0.47 &  &  &  &  &  &  & 0.42 \\
f208$^*$ & 932.994 (0.016) & 1071.818 (0.019) & 0.62 &  &  &  &  &  &  &   \\
f209 & 969.427 (0.010) & 1031.537 (0.011) & 0.27 & 2 &  &  & 4 &  & -0.08 & 0.463 \\
f210 & 969.890 (0.010) & 1031.045 (0.011) & 0.25 & 2 &  &  & 4 &  & -0.08 &   \\
f211$^{\dag}$ & 1002.650 (0.008) & 997.357 (0.008) & 0.26 &  &  &  &  &  &  & 0.272 \\
f212 & 1002.922 (0.043) & 997.087 (0.043) & 0.25 &  &  &  &  &  &  &   \\
f213 & 1012.924 (0.008) & 987.241 (0.008) & 0.24 & 9 & -9 &  &  &  &  & 0.220\\
f214 & 1013.144 (0.063) & 987.027 (0.061) & 0.18 & 9 & -8 &  &  &  &  & 0.478 \\
f215 & 1013.622 (0.030) & 986.561 (0.029) & 0.18 & 9 & -6 &  &  &  &  & 0.496 \\
f216 & 1014.118 (0.011) & 986.079 (0.011) & 0.23& 9 & -4 &  &  &  &  & 0.474 \\
f217 & 1014.592 (0.025) & 985.618 (0.024) & 0.33 & 9 & -2 &  &  &  &  & 0.195\\
t218 & 1014.787 (0.085) & 985.428 (0.083) & 0.16 & 9 & -1 & &  &  &  & 0.225\\
f219 & 1015.012 (0.035) & 985.210 (0.034) & 0.34 & 9 & 0 &  &  &  &  & 0.208 \\
f220 & 1015.220 (0.015) & 985.008 (0.015) & 0.18 & 9 & 1 &  &  &  &  & 0.243\\
f221 & 1015.463 (0.018) & 984.772 (0.017) & 0.50 & 9 & 2 &  &  &  &  & 0.417 \\
f222$^{\dag}$ & 1015.880 (0.015) & 984.368 (0.015) & 0.50 & 9 & 4 &  &  &  &  & 0.397 \\
f223 & 1016.277 (0.035) & 983.984 (0.034) & 0.74 & 9 & 6 &  &  &  &  & 0.430\\
f224 & 1016.711 (0.015) & 983.564 (0.015) & 0.45 & 9 & 8 &  &  &  &  & 0.190\\
f225 & 1016.901 (0.025) & 983.380 (0.024) & 0.58 & 9 & 9 &  &  &  &  &   \\
f226 & 1133.916 (0.006) & 881.900 (0.005) & 0.17 &  &  &  &  &  &  & 1.815 \\
f227 & 1135.731 (0.007) & 880.490 (0.005) & 0.17 &  &  &  &  &  &  &   \\
f228 & 1508.306 (0.020) & 662.995 (0.009) & 0.19 &  &  &  &  &  &  & 0.066 \\
f229 & 1508.372 (0.020) & 662.966 (0.009) & 0.22 &  &  &  &  &  &  &   \\
f230 & 1518.291 (0.010) & 658.635 (0.004) & 0.17 &  &  &  &  &  &  &   \\
f231 & 1573.084 (0.005) & 635.694 (0.002) & 0.18 &  &  &  &  &  &  &   \\
f232 & 1598.436 (0.016) & 625.612 (0.006) & 0.20 &  &  &  &  &  &  &   \\
f233 & 1650.710 (0.007) & 605.800 (0.002) & 0.23 &  &  &  &  &  &  &   \\
f234 & 1666.173 (0.007) & 600.178 (0.003) & 0.18 &  &  &  &  &  &  &   \\
f235 & 1696.132 (0.007) & 589.577 (0.002) & 0.24 &  &  &  &  &  &  &   \\
f236 & 1756.324 (0.032) & 569.371 (0.010) & 0.20 &  &  &  &  &  &  &   \\
f237 & 1846.183 (0.008) & 541.658 (0.002) & 0.18 &  &  &  &  &  &  &   \\
f238$^{\dag}$ & 2767.518 (0.008) & 361.335 (0.001) & 2.82 & 0 & 0 &  &  &  &  &   \\
f239$^{\dag}$ & 2782.454 (0.033) & 359.395 (0.004) & 0.45 & 1 & -1 &  &  &  &  & 0.747 \\
t240 & 2783.201 (0.008) & 359.299 (0.001) & 0.13 & 1 & 0 &  &  &  &  & 0.726 \\
f241 & 2783.927 (0.014) & 359.205 (0.002) & 0.20 & 1 & 1 &  &  &  &  &   \\
t242 & 2814.226 (0.037) & 355.337 (0.005) & 0.13 & 2 & -2 &  &  &  &  & 0.779 \\
t243 & 2815.005 (0.030) & 355.239 (0.004) & 0.13 & 2 & -1 &  &  &  &  & 0.73 \\
f244 & 2815.735 (0.008) & 355.147 (0.001) & 0.18 & 2 & 0 &  &  &  &  & 0.714 \\
f245 & 2816.449 (0.025) & 355.057 (0.003) & 0.65 & 2 & 1 &  &  &  &  & 0.795 \\
t246 & 2817.244 (0.031) & 354.957 (0.004) & 0.14 & 2 & 2 &  &  &  &  &   \\
f247 & 3686.925 (0.009) & 271.229 (0.001) & 0.16 &  &  &  &  &  &  & 0.422 \\
f248 & 3687.347 (0.030) & 271.198 (0.002) & 0.15 &  &  &  &  &  &  &   \\
f249 & 3700.778 (0.017) & 270.213 (0.001) & 0.17 &  &  &  &  &  &  & 1.956 \\
f250$^{\dag}$ & 3702.734 (0.088) & 270.071 (0.006) & 0.17 &  &  &  &  &  &  & 1.025 \\
f251 & 3703.759 (0.034) & 269.996 (0.002) & 0.17 &  &  &  &  &  &  & \\
\hline
\enddata
\label{tab01b}
\end{deluxetable}

\end{document}